\documentstyle[eqsecnum,aps]{revtex}
\begin{document}
\preprint{ITD 92/93--17}
\tighten
\renewcommand{\pl}{\partial}
\newcommand{\fr}{\frac} \newcommand{\noi}{\noindent}
\newcommand{\bq}{\begin{equation}} \newcommand{\eq}{\end{equation}}
\newcommand{\ra}{\rightarrow} \newcommand{\al}{\alpha}
\newcommand{\be}{\beta} \newcommand{\ga}{\gamma} \newcommand{\la}{\lambda}
\newcommand{\st}{\sqrt} \newcommand{\vph}{\varphi} \newcommand{\si}{\sigma}
\newcommand{\sla}{\;\rule[-.15cm]{.3pt}{.5cm}\; \raisebox{-.25cm}
{\footnotesize$\lambda=1$}}
\newcommand{\hf}{\frac{1}{2}} \newcommand{\iy}{\infty}
\newcommand{\res}{\left(I-K\right)^{-1}}
\newcommand{\rest}{\left(I-K^t\right)^{-1}}
\newcommand{\ac}{A}
\newcommand{\bc}{B}
\newcommand{\cc}{C}
\def\qed{\hfill \vrule height 7pt width 7pt depth 0pt \smallskip}
\def\up#1{\leavevmode \raise .3ex\hbox{$#1$}}
\def\down#1{\leavevmode \lower .5ex\hbox{$\scriptstyle#1$}}
\def\chJ{\up{\chi}_{\down{J}}}
\def\sdown#1{\leavevmode \lower .2ex\hbox{$\scriptstyle#1$}}
\title{Fredholm Determinants, Differential Equations and Matrix Models}
\author{Craig A.~Tracy\footnotemark[1]}
\footnotetext[1]{e-mail address: tracy@itd.ucdavis.edu}
\address{
Department of Mathematics and Institute of Theoretical Dynamics,\\
University of California,
Davis, CA 95616, USA}
\author{Harold Widom\footnotemark[2]}
\footnotetext[2]{e-mail address: widom@cats.ucsc.edu}
\address{
Department of Mathematics,\\
University of California,
Santa Cruz, CA 95064, USA}
\maketitle
\begin{abstract}
Orthogonal polynomial random matrix models
of $N\times N$ hermitian matrices
lead to Fredholm determinants of integral operators
with kernel of the form $\left(\vph(x) \psi(y)-\psi(x)\vph(y)\right)/x-y$.
This paper is concerned with the Fredholm determinants of integral
operators having kernel of this form and
where the underlying set is the union of intervals
$J= \bigcup_{j=1}^{m}(a_{2j-1},a_{2j})$.
The emphasis is on the determinants thought of as functions of
the end-points $a_k$.
\par
We show that these Fredholm determinants
with  kernels of the
general form described above
are expressible in terms of
solutions of systems of PDE's
 as long as $\vph$ and $\psi$ satisfy a certain type of
differentiation formula.
The $(\varphi,\psi)$ pairs for the sine, Airy, and Bessel kernels
satisfy such relations, as do the pairs which arise in the finite $N$ Hermite,
Laguerre
and Jacobi ensembles and in matrix models of 2D quantum gravity. Therefore we
shall be
able to write down the systems  of PDE's for these ensembles as special cases
of the general system.
\par
An analysis of these equations will lead  to explicit representations
in terms of Painlev{\'e} transcendents for the distribution functions
of the largest and smallest eigenvalues
 in the finite $N$ Hermite and Laguerre ensembles,
and for
 the distribution functions of the largest and smallest singular values of
rectangular matrices
 (of arbitrary dimensions) whose entries are independent identically
distributed complex Gaussian variables.
\par
There is also an exponential variant of the kernel in which the
denominator is replaced by $e^{bx}-e^{by}$, where $b$ is an arbitrary complex
number. We shall find an analogous system of differential equations in this
setting. If $b=i$
then we can interpret our operator as acting on (a subset of) the unit circle
in the complex plane. As an application of this we shall write down a system
of PDE's for Dyson's circular ensemble of $N\times N$ unitary matrices, and
then
an ODE if $J$ is an arc of the circle.
\end{abstract}
\section{Introduction}
It is a fundamental result
of Gaudin and Mehta that orthogonal polynomial random matrix models
of $N\times N$ hermitian matrices lead to integral operators
whose Fredholm determinants describe the
statistics of the spacing of eigenvalues \cite{mehta_book,porter}.
Precisely, if a weight function $w(x)$ is given, denote by $\{p_{k}(x)\}$ the
sequence
of polynomials orthonormal with respect to $w(x)$ and set
\[\vph_{k}(x):=p_{k}(x)\,w(x)^{1/2}.\]
Then $E(n;J)$, the probability that a matrix from the ensemble associated with
$w(x)$ has precisely $n$ eigenvalues in the set $J\;(n=0,1,\cdots\,)\,$, is
given by the formula
\bq
E(n;J)=\fr{(-1)^{n}}{n!}\,\fr{d^{n}}{d\la^{n}}\,\mbox{det}\,(I-\la\,K_{N})\sla
\label{detform} \eq
where $K_{N}$ is the integral operator on $J$ with kernel
\[ K_{N}(x,y):=\sum_{k=0}^{N-1}\vph_{k}(x)\,\vph_{k}(y).\]
It follows from the Christoffel-Darboux formula (c.f.~(\ref{christoffel})
below)
that $\la\,K_{N}(x,y)$ is a particular case of a kernel of the general form
\bq
 K(x,y):={\vph(x) \psi(y)-\psi(x)\vph(y)\over x-y}.
\label{kernel}\eq
\par This paper is concerned with the Fredholm determinants of integral
operators having kernel of this form and
where the underlying set is the union of intervals
\[ J:= \bigcup_{j=1}^{m}(a_{2j-1},a_{2j}).\]
The emphasis is on the determinants thought of as functions of
the end-points $a_k$.
\par
If we denote the operator itself by $K$ then it is well known that
\bq
\fr{\pl}{\pl a_k} \log \det\left(I-K\right)=(-1)^{k-1} R(a_k,a_k)
\ \ \ (k=1,\cdots,2m) \label{logdet} \eq
where $R(x,y)$, the resolvent kernel, is the kernel of $K(I-K)^{-1}$.
This requires only that $\la=1$ not be an eigenvalue of $K$ and that
$K(x,y)$ be smooth.
Jimbo, Miwa, M{\^o}ri and Sato \cite{jmms} showed for the
``sine kernel''
\[ {\sin(x-y)\over x-y}\> , \]
that if we define
\[ r_{k,\pm}:= \res e^{\pm i x}(a_k) \]
then the $R(a_k,a_k)$ are expressible in terms of the $r_{k,\pm}$
and that these in turn, as functions of the $a_1,\cdots,a_{2m}$,
satisfy a completely integrable system of partial differential
equations.  They deduced from this that in the special case
when $J$ is an interval of length $s$ the logarithmic derivative
with respect to $s$ of the Fredholm determinant satisfied a
Painlev{\'e} differential equation.  (More precisely, $s$ times this
logarithmic derivative satisfied the so-called $\sigma$ form
of $P_V$ of
Jimbo-Miwa-Okamoto \cite{jm,okamoto87}.)\ \ We refer the reader to \cite{tw1}
for a derivation of these results in  the spirit of the
present paper. The discovery that Painlev{\'e} transcendents
can be used to  represent correlation functions in statistical mechanical
models
first appeared in the 2D Ising model \cite{bmw,mtw,wmtb}.

\par
The sine kernel arises by taking a  scaling limit as $N\ra\iy$
in the bulk of the spectrum in a variety of random matrix models
of $N\times N$ hermitian matrices.
But if we take the Gaussian unitary ensemble (also called the Hermite ensemble;
see below), and others as well, and scale at the edge of the
spectrum then we are led similarly to the ``Airy kernel''
\[ { {\rm Ai}(x) {\rm Ai}'(y) - {\rm Ai}'(x) {\rm Ai}(y) \over
 x-y }\> ,\]
where ${\rm Ai}(x)$ is the Airy function
 \cite{bowick_brezin,forrester,moore,tw2}.
For this kernel the authors found \cite{tw2} a completely analogous,
although somewhat more complicated, system of PDE's, and showed
that for  $J$ a semi-infinite interval $(s,\iy)$ there was also a
Painlev{\'e} equation associated with the determinant---this time $P_{II}$.
Similarly, if we scale the Laguerre ensemble at the left edge of the
spectrum or the Jacobi ensemble at either edge (see below for these
ensembles also), then we obtain yet another kernel, the ``Bessel kernel,''
where in (\ref{kernel}) $\varphi(x)=J_\al(\sqrt{x})$,
$\psi(x)=x\varphi'(x)$ with $J_\al$ the usual Bessel function.
Again we found \cite{tw3} a system of PDE's for general $J$ and,
for $J=(0,s)$, a Painlev{\'e} equation associated with the
Fredholm determinant---this time $P_V$ (actually a special case
of $P_V$ which is reducible to $P_{III}$).
\par
In looking for (and finding) analogous systems of PDE's for finite
$N$ matrix ensembles we realized that all we needed were
differentiation formulas of a certain form for $\varphi$ and $\psi$,
namely
\begin{eqnarray}
m(x) \varphi'(x)&= A(x) \varphi(x) +  B(x) \psi(x),
\nonumber \\
m(x) \psi'(x)&= - C(x) \varphi(x)- A(x) \psi(x)
\label{diff_formulas}
\end{eqnarray}
where $m$, $A$, $B$ and $C$ are polynomials.
\par
The $(\varphi,\psi)$ pairs for the sine, Airy, and Bessel kernels
satisfy such relations ($m(x)=1$ for sine and Airy, $m(x)=x$ for
Bessel) as do the pairs which arise in the finite $N$ Hermite, Laguerre
and Jacobi ensembles ($m(x)=1$ for Hermite, $m(x)=x$ for Laguerre
and $m(x)=1-x^2$ for Jacobi) and therefore we shall be able to write
down the systems  of PDE's for these ensembles at once as special cases
of the general system.  An analysis of these equations will lead in the cases
of
the finite $N$ Hermite and Laguerre ensembles to explicit representations
in terms of Painlev{\'e} transcendents for the distribution functions
for the largest and smallest eigenvalue. A consequence of the latter
is such a representation for the distribution functions of the largest and
smallest singular values of rectangular matrices (of arbitrary dimensions)
whose entries are independent identically distributed complex Gaussian
variables;
for these singular values are the eigenvalues of a matrix from an appropriate
Laguerre ensemble \cite{edel}.

There is also an exponential variant of the kernel (\ref{kernel}) in which the
denominator is replaced by $e^{bx}-e^{by}$ (or equivalently
$\sinh\fr{b}{2}(x-y)$) where $b$ is an arbitrary complex
number. With an appropriate modification of (\ref{diff_formulas}) we shall find
a completely analogous system of differntial equations. Observe that if $b=i$
then we can interpret our operator as acting on (a subset of) the unit circle
in the complex plane. As an application of this we shall write down a system
of PDE's for Dyson's circular ensemble of $N\times N$ unitary matrices, and
then
an ODE if $J$ is an arc of the circle. In case $b$ is purely real our results
have
application to the so-called $q$-Hermite ensemble \cite{chen,muttalib}.

\par
Here, now, is a more detailed description of the contents of the paper.
%%%%%%%%%%%%%%%%%%%%%%%%%%%%%%%%%%%%%%%%%%%%%%%%%%%%%%%%%%%%%%%%%%%%%%%%%%%
\subsection{ The Differential Equations}
\label{subsec:intro_de}
In Sec.~\ref{sec:pde} we  derive our general system of partial differential
equations. To describe these equations we first
 define the functions
\bq  Q:=\res \varphi, \ \ \ P:=\res\psi, \label{PQdef}\eq
which depend also, of course, on the parameters $a_k$, and then
\begin{eqnarray}
 && q_k:=Q(a_k),\ \  p_k:=P(a_k) \ \ (k=1,\cdots,2m) \label{qpDef}  \\
u_i&:=&\left(Q(x),x^i \varphi(x)\right),\ \  v_i:=\left(Q(x),x^i\psi(x)\right),
\ \ w_i:=\left(P(x),x^i \psi(x)\right) \ \ (i=0,1,\cdots) \nonumber
\end{eqnarray}
where the inner products are taken over $J$.
These are the unknown functions in our system of PDE's.  We shall see that for
any operator with kernel of the form (\ref{kernel}) we have for the resolvent
kernel the formulas \cite{its90}
\begin{mathletters}
\begin{eqnarray}
 R(a_j,a_k)& =& {q_j p_k - p_j q_k \over a_j -a_k} \ \ \ (j\neq k),
\label{Rjk}\\
R(a_k,a_k)&=& p_k \fr{\pl q_k}{\pl a_k} - q_k\fr{\pl p_k}{\pl a_k}\> ,
\label{Rkk_intro}
\end{eqnarray}
\end{mathletters}
for the $q_j$ and $p_j$ the differentiation formulas
\begin{eqnarray}
\fr{\pl q_j}{\pl a_k}&=& (-1)^k R(a_j,a_k) q_k \ \ \ (j\neq k),
\label{qjk} \\
\fr{\pl p_j}{\pl a_k}&=& (-1)^k R(a_j,a_k) p_k \ \ \ (j\neq k),
\label{pjk}\end{eqnarray}
and for the $u_j$, $v_j$, $w_j$ differentiation formulas of the form
\[ \fr{\pl u_j}{\pl a_k},\  \fr{\pl v_j}{\pl a_k},\  \fr{\pl w_j}{\pl a_k} =
{\rm polynomial\  in }\ p_k, q_k \ {\rm and\  the\  various }\ u_i, v_i, w_i.
\]
These equations are universal for any kernel of the form (\ref{kernel}).
What depends on (\ref{diff_formulas})
are the remaining differential formulas
\begin{eqnarray*}
m(a_j) \fr{\pl q_j}{\pl a_j}& = &{\rm polynomial\ in}\ q_j,\ p_j\ {\rm and\
the}\
 u_i,\ v_i,\ w_i\\
&&- \sum_{k\neq j} (-1)^k R(a_j,a_k) q_k, \\
m(a_j) \fr{\pl p_j}{\pl a_j}& = &{\rm polynomial\ in}\ q_j,\ p_j\ {\rm and\
the}\
 u_i,\ v_i,\ w_i\\
&&- \sum_{k\neq j} (-1)^k R(a_j,a_k) p_k\, ,
\end{eqnarray*}
and the representation
\[ m(a_j) R(a_j,a_j) = {\rm polynomial\ in}\ q_j,\ p_j\ {\rm and\ the}\
 u_i,\ v_i,\ w_i .\]
The polynomials on the right sides  are expressed in terms of the coefficients
of the polynomials $m$, $\ac$, $\bc$, $\cc$ in (\ref{diff_formulas}).
We mention that in \cite{jmms} no ``extra'' quantities $u_i$, $v_i$, $w_i$
appear, but this is quite special.  In general the number of triples
$(u_i,v_i,w_i)$ which occur is at  most
\[ \max\left(\deg\ac,\deg\bc,\deg\cc,\deg m-1\right), \]
although in practice fewer of these quantities actually appear.

%%%%%%%%%%%%%%%%%%%%%%%%%%%%%%%%%%%%%%%%%%%%%%%%%%%%%%%%%%%%%%%%%%%%%%%%%%
\subsection{The Examples}
First in Sec.~\ref{sec:sine_airy_bessel} we quickly derive, as special cases,
the systems of equations for the sine, Airy and Bessel kernels.  Then in
Sec.~\ref{sec:beyond_airy} we derive and investigate the equations for
kernels ``beyond Airy.''
To explain this we replace the variables $x$, $y$ in (\ref{kernel}) by
$\la$ and $\mu$, think of (a completely new variable) $x$ as a parameter,
and observe that for each $x$
\bq
 K(\la,\mu):={{\rm Ai}(x+\la) {\rm Ai}'(x+\mu) -
{\rm Ai}'(x+\la) {\rm Ai}(x+\mu) \over \la - \mu}
\label{general_airy}\eq
has the same properties as the Airy kernel. (In the differentiation
formulas (\ref{diff_formulas}) the variable is now $\la$ and $x$ is a parameter
in the coefficients.)\ \ Observe also that ${\rm Ai}(x+\la)$ is, as a function
of $x$, an eigenfunction of the Schr{\"o}dinger operator with
potential $\xi(x)=-x$ corresponding to eigenvalue $\la$.
\par
In the hermitian matrix models of 2D quantum gravity
\cite{brezin_kazakov,brezin_review,douglas_shenker,douglas,gross_migdal}
solutions to the so-called string equation
\[ \left[{\cal Q},{\cal P}\right]=1 \]
determine the functions $\varphi$ and $\psi$.  In the simplest case of
the KdV hierarchy,  the operator ${\cal Q}$ is the Schr{\"o}dinger operator
(note our convention of sign of $D_{x}^{2}$)
\[ {\cal Q} = D_x^2 + \xi(x) \]
and the differential operator   ${\cal P}$ (in $x$)
is
\[ {\cal P} = \left({\cal Q}^{(2\ell -1)/2}\right)_+ \ \ \ (\ell=1,2,\cdots) \]
where $(\cdot)_+$ is the differential operator part.  The potential $\xi$
then satisfies a differential equation determined
by the string equation and  $\varphi(\la,x)$ is
the  eigenfunction
\bq
 \left(D_x^2 + \xi(x)\right)\varphi(\la,x) = \la \varphi(\la,x)
\label{schrodinger}
\eq
satisfying
\bq
 \fr{\pl\varphi}{\pl\la} = {\cal P}\varphi(\la,x). \label{P_operator}\eq
Setting
\bq \psi(\la,x)=D_x\varphi(\la,x) \label{psi_beyond_airy}\eq
the kernel \cite{bowick_brezin,moore} is then
\begin{eqnarray}
 K(\la,\mu)& =&{\varphi(\la,x)\psi(\mu,x)-\psi(\la,x)\varphi(\mu,x)\over
\la-\mu}\nonumber \\
&=& \int_x^\iy \varphi(\la,y)\varphi(\mu,y)\, dy
 \, . \label{kernel_double}\end{eqnarray}
These are the kernels which we say are ``beyond Airy'' since
for $\ell=1$, ${\cal P}=D_x$, $\xi(x)=-x$,  (\ref{kernel_double}) reduces to
the generalized Airy kernel (\ref{general_airy}). From (\ref{schrodinger})
and (\ref{P_operator}) it follows that for general $\ell$ the functions
$\varphi(\la,x)$ and $\psi(\la,x)$ satisfy differentiation formulas (in $\la$)
of the form (\ref{diff_formulas}).  (Again in the differentiation formulas
(\ref{diff_formulas}) the variable is now $\la$ and $x$ is a parameter in
the coefficients.) \ \  In Sec.~\ref{sec:beyond_airy} we illustrate these
methods for the case $\ell=2$.
\par
In Sec.~\ref{sec:her_lag_jac}  we study in some detail the finite $N$
Hermite, Laguerre, Jacobi, and circular ensembles.  In orthogonal polynomial
ensembles one is given a weight function $w(x)$ and then, for any symmetric
function $f$ on ${\bf R}^N$,  we have
\bq E\left(f(\la_1,\cdots,\la_N)\right)= c_N \int\cdots\int f(x_1,\cdots,x_N)
\prod w(x_i) \prod_{i< j}\left\vert x_i-x_j\right\vert^2\, dx_1\cdots dx_N
\label{expected} \eq
where ``$E$'' denotes the expected value, $\la_1,\cdots,\la_N$ are the
eigenvalues, and $c_N$ is a constant such that the right side equals one
when $f\equiv 1$.  In the Hermite ensemble $w(x)=e^{-x^2}$ and the integrations
are over ${\bf R}$, in the Laguerre ensemble $w(x)=x^\al e^{-x}$ and the
integrations are over ${\bf R}^+$, and in the Jacobi ensemble
$w(x)=(1-x)^\al(1+x)^\be$ and the integrations are over $(-1,1)$.  In the
circular ensemble $w(x)=1$ and the integrations are over the unit
circle.
\par
The size parameter $N$ will appear only as a coefficient parameter in the
equations we obtain; and we find that the equations for both bulk
and edge scaling limits emerge as limiting cases.  Our equations also make
the study of large $N$ corrections to the scaling limits tractable.
\par
For the Hermite, Laguerre and Jacobi ensembles there are natural intervals
depending upon a single parameter $s$---for Hermite $J=(s,\iy)$ or
$(-\iy,s)$, for Laguerre $J=(s,\iy)$ or $(0,s)$, and for Jacobi
$J=(s,1)$ or $J=(-1,s)$---and in all these
cases we shall find an associated ordinary differential equation. For Hermite
and
Laguerre these will be of Painlev\'{e} type. Observe that taking $n=0$ in
(\ref{detform}) shows that
the Fredholm determinant in each of these cases is precisely the
distribution function for the largest eigenvalue, or 1 minus the distribution
for the smallest eigenvalue.
\subsection{General Matrix Ensembles}
In this  final section of the paper we show that there are differentiation
formulas of the form (\ref{diff_formulas}) when Hermite, Laguerre, or
Jacobi weights are multiplied by $e^{-V(x)}$ where $V(x)$ is an
arbitrary polynomial.  (Of course it must be of such a form that the resulting
integrals are convergent.)\ \ It is the finite $N$ matrix models corresponding
to certain $V(x)$ which, in an appropriate double scaling limit at the
edge of the spectrum, lead to the kernels beyond Airy.  (Strictly speaking
this is true only for the universality classes $\ell=1,3,5,\ldots$ as it is
well-known that the cases $\ell=2,4,6,\ldots$ require coefficients in $V(x)$
that make $e^{-V(x)}$ unbounded.)
%%%%%%%%%%%%%%%%%%%%%%%%%%%%%%%%%%%%%%%%%%%%%%%%%%%%%%%%%%%%%%%%%%%%%
\section{The General System of Partial Differential Equations}
\label{sec:pde}
In this section we derive the system of partial differential
equations
that determine the functional dependence of the Fredholm determinant
$\det(I-K)$  upon the parameters $a_k$ where $K$ has kernel
(\ref{kernel}).
 After
some  preliminary definitions and
identities  in Sec.~\ref{subsec:preliminary},
in Sec.~\ref{subsec:universal} we derive those
equations which are independent of the differentiation formulas
(\ref{diff_formulas}).  In Sec.~\ref{subsec:m=1} we assume
$\varphi$ and $\psi$ satisfy the differentiation formulas
for the case  $m(x)=1$.  Then
in Sec.~\ref{subsec:m} we indicate
the modifications necessary
 for the general case of polynomial $m$.
Finally, in Sec.~\ref{subsec:DE_circle} we derive the exponential variant
of the system of equations.
\par
\subsection{Preliminaries}
\label{subsec:preliminary}
Our derivation will use, several times, the commutator identity
\bq [L,(I-K)^{-1}]\,=\,(I-K)^{-1}[L,K](I-K)^{-1}, \label{comid} \eq
which holds for arbitrary operators $K$ and $L$, and the differentiation
formula
\bq\frac{d}{da}\,\res\,=\,(I-K)^{-1}\,\frac{dK}{da}\,(I-K)^{-1},\label{dform}
\eq
which holds for an arbitrary operator depending smoothly on a parameter $a$.
\par
It will be convenient to think of our operator $K$ as acting, not on $J$,
but on a larger natural domain ${\cal D}$ and to have kernel
\bq  K(x,y)\chJ(y) \label{kernel_chJ} \eq
where $\chJ$ is the characteristic function of $J$
and $K(x,y)$ is the kernel (\ref{kernel}).  For example, for
the sine and Airy kernel  ${\cal D}={\bf R}$, for the Bessel kernel
${\cal D}={\bf R}^+$, and for the Jacobi
kernel  ${\cal D}=(-1,1)$.  The set $J$ will
 be  a subset of ${\cal D}$.
We will continue to denote the resolvent kernel of $K$ by $R(x,y)$ and
note that it is smooth in $x$ but discontinuous at $y=a_k$.
We will also  need the distributional kernel
\[ \rho(x,y)=\delta(x-y)+R(x,y)\]
 of $\res$.   The quantities
$R(a_j,a_k)$ in Sec.~\ref{subsec:intro_de}  are interpreted to mean
\[\lim_{\stackrel{y\ra a_{k}}{y\in \sdown{J}}}R(a_{j},y),\]
and similarly for $p_{j}$ and $q_{j}$.
\par
The definitions of $u_i$, etc.\ must be modified.  Before doing this
it will be  convenient to introduce
\bq Q_j(x):=\res x^j\varphi(x),\ \ P_j(x):=\res x^j \psi(x) \ \ \
 \label{PQjDef}\eq
which for $j=0$ reduce to (\ref{PQdef}) ($Q_0=Q$, $P_0=P$).
Then we define
\begin{mathletters}
\begin{eqnarray}
u_j:=\left(Q,x^j\varphi\chJ\right)=\left(Q_j,\varphi\chJ\right), \label{ujDef}
\\
v_j:=\left(Q,x^j\psi\chJ\right)=\left(P_j,\varphi\chJ\right), \label{vjDef}\\
\tilde v_j:=\left(P,x^j\varphi\chJ\right)=
\left(Q_j,\psi\chJ\right),\label{vtildejDef} \\
w_j:=\left(P,x^j\psi\chJ\right)=\left(P_j,\psi\chJ\right)
\label{wkDef}
\end{eqnarray}
\end{mathletters}
where the inner product $(\cdot,\cdot)$  is now over the  domain ${\cal D}$.
That these definitions
for $u_j$, $v_j$ and $w_j$
 are equivalent to those of Sec.~\ref{subsec:intro_de}
and the above equalities are left as exercises.  They follow from
 the fact that
\bq
 \rest f\chJ =\res f\chJ = \left \{ \begin{array}{l}
	\res f \;\; \mbox{ on $J$} \\
	0 \;\; \mbox{on $J^c$}\end{array} \right.
\label{identity1}\eq
for any smooth $f$.
Here $K^t$ is the transpose of the operator $K$. (Note that $K$ takes
smooth functions to smooth functions while its transpose takes distributions
to distributions.)\ \ We also observe that $\tilde v_0=v_0$.
A final bit of preliminary notation is
\[ L\doteq L(x,y)\]
which means the operator $L$ has kernel $L(x,y)$.
%%%%%%%%%%%%%%%%%%%%%%%%%%%%%%%%%%%%%%%%%%%%%%%%%%%%%%%%%%%%%%%%%%%%%%%%%%%
\subsection{The Universal Equations}
\label{subsec:universal}
In this subsection $M$ denotes multiplication by the independent variable.
If we consider the commutator of $M$ with $K$ and use the representation
(\ref{kernel_chJ}), we have immediately
\[ \left[M,K\right]\doteq\left(\varphi(x)\psi(y)-\psi(x)\varphi(y)\right)
\chJ(y),  \]
and so by (\ref{comid})
\bq
 \left[M,\res\right]\doteq Q(x)\rest \psi\chJ(y) - P(x)\rest\varphi\chJ(y).
\label{M_com1}
\eq
(The transpose here arises from the general fact that if $L\doteq U(x)V(y)$
then
$T_{1}LT_{2}\doteq T_{1}U(x)\,T_{2}^{t}V(y)$.)\ \ It follows immediately that
\[ \left[M,\res\right]\doteq (x-y)\rho(x,y)=(x-y)R(x,y),\]
and hence referring to (\ref{identity1}), we deduce
\[ R(x,y)={Q(x)P(y)-P(x)Q(y)\over x-y} \ \ \ (x,y\in J). \]
In particular we have deduced (\ref{Rjk})
(recall definitions (\ref{qpDef})) and the representation
\bq R(x,x)\,=\,Q'(x)\,P(x)-P'(x)\,Q(x) \hspace{3em}(x\in J). \label{Rxx} \eq
We remark that the generality of
this   expression for $R(x,y)$ was first, as far
as the authors are aware, stressed by Its, et.~al.~\cite{its90} though
it appears, of course,  in the
context of the sine kernel in the earlier work of JMMS \cite{jmms}.
\par
We have the easy fact that
\[ \fr{\pl}{\pl a_{k}}K\doteq(-1)^{k}K(x,a_{k})\,\delta(y-a_{k})\]
and so by (\ref{dform})
\bq \fr{\pl}{\pl a_{k}}(I-K)^{-1}\doteq(-1)^{k}R(x,a_{k})\,\rho(y,a_{k}).
\label{resder} \eq
At this point we use the notations $Q(x,a),\, P(x,a)$ for $P(x)$ and $Q(x)$,
respectively, to remind
ourselves that they are functions of $a$ as well as $x$. We deduce immediately
from (\ref{resder}) and (\ref{PQdef}) that
\bq \fr{\pl}{\pl a_{k}}Q(x,a)=(-1)^{k}R(x,a_{k})q_{k},\;\;\fr{\pl}{\pl a_{k}}
P(x,a)=(-1)^{k}R(x,a_{k})p_{k}. \label{PQpls} \eq
Since $q_{j}=Q(a_{j},a)$ and $p_{j}=P(a_{j},a)$ this gives
\[\fr{\pl q_{j}}{\pl a_{k}}=(-1)^{k}R(a_{j},a_{k})q_{k},\hspace{1em}
\fr{\pl p_{j}}{\pl a_{k}}=(-1)^{k}R(a_{j},a_{k})p_{k},\hspace{2em}(j\neq k).\]
These are equations (\ref{qjk}) and (\ref{pjk}).
We record for use below
\bq\fr{\pl q_{j}}{\pl a_{j}}=(\fr{\pl}{\pl x}+\fr{\pl}{\pl a_{j}})
Q(x,a)\Bigr\vert_{x=a_{j}},
\hspace{1em}\fr{\pl p_{j}}{\pl a_{j}}=(\fr{\pl}{\pl x}+\fr{\pl}{\pl a_{j}})
P(x,a_{j})\Bigr\vert_{x=a_{j}}.\label{chain_rule}\eq

To obtain (\ref{Rkk_intro}) observe that (\ref{Rxx}) gives
\[ R(a_k,a_k)=\fr{\pl}{\pl x}\left[ Q(x,a) p_k - P(x,a)
q_k\right]\bigg\vert_{x=a_k}
\, . \]
But the expression in brackets above vanishes identically when $x=a_k$ and
so the above is equal to
\[ - \fr{\pl}{\pl a_k}\left[ Q(x,a) p_k - P (x,a) q_k
\right]\bigg\vert_{x=a_k}\> . \]
If we use (\ref{PQpls}) in the computation of this partial derivative,
(\ref{Rkk_intro}) results.
 \par
We now show that the $\tilde v_k$ can be expressed in terms of the
other quantities $u_i$, $v_i$ and $w_i$ (we could do $v_k$ just as
well) and that the $P_k$ and $Q_k$ can be expressed in terms of these
quantities and $P$, $Q$.  From
\[ \fr{x^k-y^k}{x-y}=\sum_{\stackrel{i+j=k-1}{i,j\geq 0}} x^i y^j \]
we get
\[ \left[M^k,K\right]\doteq \sum_{\stackrel{i+j=k-1}{i,j\geq 0}}
\left( x^i\varphi(x)\, y^j\psi(y) - x^i\psi(x)\, y^j\varphi(y)\right)\chJ(y)
\]
and so
 \[ \left[M^k,\res\right]\doteq\sum_{\stackrel{i+j=k-1}{i,j\geq 0}}
\left(Q_i(x)\,\rest y^j\psi(y)\chJ(y)-P_i(x)\,\rest y^j\varphi(y)\chJ(y)
\right).\]
Applying this to $\varphi$ shows that
\bq
Q_k(x)=x^k Q(x) - \sum_{\stackrel{i+j=k-1}{i,j\geq 0}}
\left( v_j Q_i(x) - u_j P_i(x)\right)\label{Qk_recursion}\eq
and applying it to $\psi$ shows that
\bq
P_k(x)=x^k P(x)-\sum_{\stackrel{i+j=k-1}{i,j\geq 0}}
\left(w_j Q_i(x) -\tilde v_j P_i(x)\right).\label{Pk_recursion}\eq
These are the recursion relations for $Q_k$, $P_k$.  Taking
the inner product of both sides of the first one with $\psi \chJ$
gives
\bq \tilde v_k=v_k-\sum_{\stackrel{i+j=k-1}{i,j\geq 0}}
\left(v_j \tilde v_i - u_j w_i\right),\label{vktilde_recursion}\eq
recursion formulas which can be used to express the $\tilde v_k$
in terms of the $u_i$, $v_i$, $w_i$.
\par
Finally, using the definition of $u_j$ in (\ref{ujDef}), the fact
\[ \fr{\pl}{\pl a_k} \chJ(y)=(-1)^k\delta(y-a_k) \, ,\]
and (\ref{resder}) we find that
\begin{eqnarray}
 \fr{\pl u_j}{\pl a_k}&=& (-1)^k\left(R(x,a_k),x^j \varphi\chJ\right)
+ (-1)^k Q(a_k) a_k^j \varphi(a_k)\nonumber\\
&=&(-1)^k\left(\rho(x,a_k),x^j\varphi\chJ\right) q_k \nonumber\\
&=&(-1)^k Q_j(a_k) q_k\nonumber\\
&=&(-1)^k q_{jk}\, q_k \label{uj_eqn}\end{eqnarray}
where $q_{jk}=Q_j(a_k)$ ($q_{0k}=q_k$).
Similarly,
\begin{eqnarray}
\fr{\pl v_j}{\pl a_k}&=(-1)^k P_j(a_k) Q(a_k)=(-1)^k p_{jk}\, q_k
\label{vj_eqn}
\\
\fr{\pl \tilde v_j}{\pl a_k}&=(-1)^k Q_j(a_k) P(a_k)= (-1)^k q_{jk}\, p_k
\label{vtildej_eqn}\\
\fr{\pl w_j}{\pl a_k}&=(-1)^k P_j(a_k) P(a_k)=(-1)^k p_{jk}\,  p_k
\label{wj_eqn}
\end{eqnarray}
where $p_{jk}=P_j(a_k)$ ($p_{0k}=p_k$).
{}From (\ref{Qk_recursion}), (\ref{Pk_recursion})
and (\ref{vktilde_recursion})  we recall that
$q_{jk}$ and $p_{jk}$ are expressible in terms of the $q_j$, $p_j$,
$u_i$, $v_i$, and $w_i$.

\subsection{ The Case $m(x)=1$}
\label{subsec:m=1}
In this section we derive those partial differential equations
that depend upon the differentiation formulas (\ref{diff_formulas})
in the special case  $m(x)\equiv 1$. We let $D$ denote the
differentiation operator with respect to the independent variable
and recall that if the operator $L$ has distributional kernel
$L(x,y)$ then
\bq  \left[D,L\right]\doteq \left(\fr{\pl}{\pl x}+\fr{\pl}{\pl y}\right)
L(x,y).
\label{DL_com}\eq
\par
Using the differentiation formulas it follows that
\begin{eqnarray}
\FL
\left(\fr{\pl}{\pl x}+\fr{\pl}{\pl y}\right)K(x,y)&=&
\fr{\ac(x)-\ac(y)}{x-y}\left(\varphi(x)\psi(y)+\psi(x)\varphi(y)\right)
\nonumber \\
&& + {\bc(x)-\bc(y)\over x-y}\psi(x)\psi(y)
+{\cc(x)-\cc(y)\over x-y}\varphi(x)\varphi(y) .
\label{K_deriv}\end{eqnarray}
Let us write
\bq
\ac(x)=\sum \al_j x^j, \ \ \bc(x)=\sum \be_j x^j, \ \ \cc(x)=\sum\ga_j x^j.
\label{ABC_polys}
\eq
Then
\[ \fr{\ac(x)-\ac(y)}{x-y}= \sum_{j,k\geq 0} \al_{j+k+1} x^j y^k, \  {\rm etc.}
\]
and we obtain the identity (recall (\ref{kernel_chJ}) and (\ref{DL_com}))
\begin{eqnarray*}
[D,K] & \doteq & \sum_{j,k\geq 0} \al_{j+k+1}
 \left( x^j \varphi(x) y^k \psi(y)
+x^j\psi(x) y^k \varphi(y)\right) \chJ(y)
 + \sum_{j,k\geq 0} \be_{j+k+1} x^j \psi(x) y^k \psi(y) \chJ(y) \\
&& + \sum_{j,k\geq 0} \ga_{j+k+1} x^j \varphi(x) y^k \varphi(y) \chJ(y)
- \sum_k (-1)^k K(x,a_k) \delta(y-a_k)
\end{eqnarray*}
from which it follows that
\begin{eqnarray}
\left[D,\res\right]&\doteq&
\sum\al_{j+k+1}\left(Q_j(x)\rest y^k\psi(y)\chJ(y)
+ P_j(x)\rest y^k \varphi(y)\chJ(y)\right)\nonumber\\
&&+\sum\be_{j+k+1} P_j(x)\rest y^k \psi(y)\chJ(y)\nonumber\\
&&+\sum\ga_{j+k+1} Q_j(x) \rest y^k \varphi(y) \chJ(y)\nonumber\\
&&-\sum(-1)^k R(x,a_k) \rho(a_k,y).
\label{Dres_com1}\end{eqnarray}
\par
We now use this last commutator to compute $Q'(x)$ and $P'(x)$:
\begin{eqnarray*}
Q'(x)&=& D\res \varphi(x) \nonumber\\
&=& \res D\varphi(x) + \left[D,\res\right]\varphi(x)\nonumber\\
&=& \res D\varphi(x) + \sum\al_{j+k+1}\left(v_k Q_j(x)+u_k P_j(x)\right)
\nonumber\\
&&+\sum\be_{j+k+1} v_k P_j(x) + \sum \ga_{j+k+1} u_k Q_j(x) \nonumber \\
&& - \sum (-1)^k R(x,a_k) q_k
\end{eqnarray*}
and similarly
\begin{eqnarray*}
P'(x)&=&\res D\psi(x)+\sum\al_{j+k+1}\left(w_k Q_j(x)+\tilde v_k P_j(x)\right)
\nonumber\\
&&+\sum\be_{j+k+1} w_k P_j(x)+\sum\ga_{j+k+1} \tilde v_k Q_j(x)
\nonumber \\
&& - \sum (-1)^k R(x,a_k) p_k.
\end{eqnarray*}
Finally we use the differentiation formulas (\ref{diff_formulas}) and
representations (\ref{ABC_polys}) to deduce
\begin{eqnarray*}
\res D\varphi(x)&=& \sum\left(\al_j Q_j(x)+\be_j P_j(x)\right), \\
\res D\psi(x)&=&\sum\left(-\ga_j Q_j(x)-\al_j P_j(x)\right)
\end{eqnarray*}
and so substituting into the above gives
\begin{eqnarray}
 Q'(x)&=&\sum_{j\geq 0}\left(\al_j+\sum_{k\geq 0} \al_{j+k+1} v_k
+\sum_{k\geq 0}\ga_{j+k+1} u_k\right) Q_j(x) \nonumber\\
&&+\sum_{j\geq 0}\left(\be_j+\sum_{k\geq 0}\al_{j+k+1}u_k+
\sum_{k\geq 0}\be_{j+k+1} v_k\right) P_j(x) \nonumber\\
&&-\sum_{k=1}^{2m} (-1)^k R(x,a_k) q_k,\label{Q'}\\
\vspace{8pt}
P'(x)&=&\sum_{j\geq 0}\left(-\ga_j+\sum_{k\geq 0}\al_{j+k+1} w_k
+\sum_{k\geq 0}\ga_{j+k+1} \tilde v_k\right) Q_j(x) \nonumber\\
&&+\sum_{j\geq 0}\left(-\al_j+\sum_{k\geq 0}\al_{j+k+1}\tilde v_k +
\sum_{k\geq 0}\be_{j+k+1} w_k\right) P_j(x) \nonumber\\
&&-\sum_{k=1}^{2m} (-1)^k R(x,a_k) p_k.\label{P'}
\end{eqnarray}
{}From (\ref{PQpls}), (\ref{chain_rule}) and these last identities we
deduce the equations
\begin{eqnarray}
\fr{\pl q_i}{\pl a_i}&=&\sum_{j\geq 0}\left(\al_j+\sum_{k\geq 0}
 \al_{j+k+1} v_k
+\sum_{k\geq 0}\ga_{j+k+1} u_k\right) q_{ji}\nonumber \\
&&+\sum_{j\geq 0}\left(\be_j+\sum_{k\geq 0}\al_{j+k+1}u_k+
\sum_{k\geq 0}\be_{j+k+1} v_k\right) p_{ji} \nonumber \\
&&-\sum_{k\neq i} (-1)^k R(a_i,a_k) q_k,\label{qi_ai}\\
\vspace{8pt}
\fr{\pl p_i}{\pl a_i}&=&\sum_{j\geq 0}\left(-\ga_j+
\sum_{k\geq 0}\al_{j+k+1} w_k
+\sum_{k\geq 0}\ga_{j+k+1} \tilde v_k\right) q_{ji}\nonumber \\
&&+\sum_{j\geq 0}\left(-\al_j+\sum_{k\geq 0}\al_{j+k+1}\tilde v_k +
\sum_{k\geq 0}\be_{j+k+1} w_k\right) p_{ji}\nonumber \\
&&-\sum_{k\neq i}(-1)^k R(a_i,a_k) p_k. \label{pi_ai}
\end{eqnarray}
Using (\ref{Rxx}), (\ref{Q'}), and (\ref{P'}) we deduce
\begin{eqnarray}
R(a_i,a_i)&=&\sum_{j\geq 0}\left(\al_j+\sum_{k\geq 0} \al_{j+k+1} v_k
+\sum_{k\geq 0}\ga_{j+k+1} u_k\right) q_{ji}\, p_i \nonumber\\
&&+\sum_{j\geq 0}\left(\be_j+\sum_{k\geq 0}\al_{j+k+1}u_k+
\sum_{k\geq 0}\be_{j+k+1} v_k\right) p_{ji}\, p_i \nonumber\\
&&+\sum_{j\geq 0}\left(\ga_j-\sum_{k\geq 0}\al_{j+k+1} w_k
-\sum_{k\geq 0}\ga_{j+k+1} \tilde v_k\right) q_{ji}\,  q_i \nonumber\\
&&+\sum_{j\geq 0}\left(\al_j-\sum_{k\geq 0}\al_{j+k+1}\tilde v_k-
\sum_{k\geq 0}\be_{j+k+1} w_k\right) p_{ji}\,  q_i \nonumber\\
&&+\sum_{k\neq i} (-1)^k \fr{\left(q_ip_k-p_iq_k\right)^2}{a_i-a_k}\; .
\label{Rii}
\end{eqnarray}

We end this section with two differentiation formulas for $R(a_i,a_i)$. From
(\ref{resder}) and (\ref{Dres_com1}) we deduce that for $x,y\in J$
\[\left( \fr{\pl}{\pl a_i}+ \fr{\pl}{\pl x}+\fr{\pl}{\pl y}\right) R(x,y)
=\sum\al_{j+k+1}\left( Q_{j}(x)P_{k}(y)+P_{j}(x)Q_{k}(y)\right)\]
\[+\sum\be_{j+k+1}P_{j}(x)P_{k}(y)+\sum\ga_{j+k+1}Q_{j}(x)Q_{k}(y)
-\sum_{k\neq i}(-1)^{k}R(x,a_{k})R(a_{k},y).\]
Hence, using the chain rule,
\begin{eqnarray}
\fr{\pl}{\pl a_i}R(a_{i},a_{i})&=&2\sum_{j,k\geq 0}\al_{j+k+1}\,q_{ji}\,p_{ki}+
\sum_{j,k\geq 0}\be_{j+k+1}\,p_{ji}\,p_{ki}
\nonumber\\
&& +\sum_{j,k\geq 0}\ga_{j+k+1}\,q_{ji}\,q_{ki}-
\sum_{k\neq i}(-1)^{k}R(a_{i},a_{k})^{2}.
\label{Riidform} \end{eqnarray}
A variant of this follows from it and (\ref{resder}) by the chain rule:
\begin{eqnarray}
\fr{d}{dt}R_{tJ}(t\,a_{i},t\,a_{i})&=&2a_{i}\sum_{j,k\geq 0}\al_{j+k+1}
\,q_{ji}\,p_{ki}+
a_{i}\sum_{j,k\geq 0}\be_{j+k+1}\,p_{ji}\,p_{ki}
\nonumber\\
&& +a_{i}\sum_{j,k\geq 0} \ga_{j+k+1}\,q_{ji}\,q_{ki}
+\sum_{k\neq i}(-1)^{k}(a_{k}-a_{i})R_{tJ}(t a_{i},t a_{k})^{2}.
\label{GG}\end{eqnarray}
Here the subscripts $tJ$ indicate that this is the underlying interval.
\subsection{The Case of Polynomial $m$}
\label{subsec:m}
Now let us see how the above derivation has to be modified
if $m(x)$ is an arbitrary polynomial.  In this section $M$ denotes
multiplication by $m(x)$ and $D$ continues to denote differentiation
with respect to the independent variable. In place of the commutator
$\left[D,K\right]$  we consider the commutator
\begin{eqnarray*}
\left[MD,K\right]&\doteq& \left(m(x)\fr{\pl}{\pl x}+m(y)\fr{\pl}{\pl y}
+m'(y)\right)K(x,y)\chJ(y)\\
&& -\sum (-1)^k m(a_k) K(x,a_k) \delta(y-a_k)\end{eqnarray*}
while using (\ref{diff_formulas}) we compute that
\bq
\left(m(x)\fr{\pl}{\pl x}+ m(y)\fr{\pl}{\pl y}+
\fr{m(x)-m(y)}{x-y}
\right)K(x,y)=\mbox{ the right hand side of (\ref{K_deriv})}.
\label{KM_deriv}\eq
Therefore if $m(x)$ is linear and we replace $D$ by $MD$ on the left side of
(\ref{Dres_com1}), then the right side
has to be changed only by insertion of factors $m(a_k)$ in the last
summand.  It follows from this that (\ref{qi_ai}) and (\ref{pi_ai})
require only the following modifications:
\begin{eqnarray}
\centerline{
\mbox{ Insert on the left sides of (\ref{qi_ai}) and (\ref{pi_ai})
the factor $m(a_i)$.}}\nonumber\\
\centerline{\mbox{ Insert in the last summands on the right sides of
(\ref{qi_ai}) and (\ref{pi_ai})}}\nonumber\\
\centerline{\mbox{ the factors $m(a_k)$.}}
\label{m_insertions1}
\end{eqnarray}
while (\ref{Rii}) and (\ref{Riidform}) require the following:
\begin{eqnarray}
\centerline{
\mbox{ Insert on the left sides of (\ref{Rii}) and (\ref{Riidform})
the factor $m(a_{i})$ in front of $R(a_{i},a_{i})$.}}\nonumber\\
\centerline{\mbox{ Insert in the last summands of (\ref{Rii}) and
(\ref{Riidform}) the factors $m(a_{k})$.}}
\label{mR_insertions1}\end{eqnarray}
For general $m(x)$, if
\[ m(x)=\sum\mu_k x^k \]
then
\[
\fr{1}{x-y}\left( m'(y)-\fr{m(x)-m(y)}{x-y}\right)
=-\sum_{\stackrel{j,k\geq 0}{j+k\leq \deg m -2}} (k+1) \mu_{j+k+2}\,  x^j y^k\,
 . \]
It follows from this and (\ref{KM_deriv}) that
\begin{eqnarray*}
\left(m(x)\fr{\pl}{\pl x}+ m(y)\fr{\pl}{\pl y}+
m'(y)
\right)K(x,y)&=&\mbox{ the right hand side of (\ref{K_deriv})}\\
&&
-\sum (k+1) \mu_{j+k+2}\, x^j y^k \left(\varphi(x)\psi(y)-\psi(x)\varphi(y)
\right).
\end{eqnarray*}
So for $m(x)$ of degree greater than $1$ the right sides of (\ref{Q'})
and (\ref{P'}) must also be modified by the addition, respectively,
of the terms
\begin{eqnarray*}
-\sum (k+1)\mu_{j+k+2} \left(v_k Q_j(x)-u_k P_j(x)\right),  \\
-\sum (k+1)\mu_{j+k+2} \left(w_k Q_j(x) - \tilde v_k P_j(x) \right ).
\end{eqnarray*}
The upshot is that in this general case (\ref{qi_ai}) and (\ref{pi_ai})
require, in addition to  (\ref{m_insertions1}), the following
modifications
\begin{eqnarray}
\mbox{Add to the right side of (\ref{qi_ai})}\nonumber\\
-\sum_{\stackrel{j,k\geq 0}{j+k\leq \deg m -2}}
(k+1)\mu_{j+k+2} \left(v_k q_{ji}-u_k p_{ji} \right).\nonumber \\
\mbox{Add to the right side of (\ref{pi_ai})} \nonumber \\
-\sum_{\stackrel{j,k\geq 0}{j+k\leq \deg m -2}}
(k+1)\mu_{j+k+2} \left(w_k q_{ji} - \tilde v_k p_{ji} \right ).
\label{m_insertions2}
\end{eqnarray}
And for general $m(x)$ we must modify (\ref{Rii}) and (\ref{Riidform}),
in addition to (\ref{mR_insertions1}), by the following:
\begin{eqnarray}
\mbox{Add to the right side of (\ref{Rii})}\nonumber \\
-\sum_{\stackrel{j,k\geq 0}{j+k\leq \deg m -2}}
 (k+1)\mu_{j+k+2} \left(v_k q_{ji}-u_k p_{ji} \right)p_i \nonumber \\
+\sum_{\stackrel{j,k\geq 0}{j+k\leq \deg m -2}}
 (k+1)\mu_{j+k+2} \left(w_k q_{ji} - \tilde v_k p_{ji} \right )q_i\, .
\label{mR_insertions2}
\end{eqnarray}
\bq
\mbox{Add to the right side of (\ref{Riidform})}\;
\sum_{\stackrel{j,k\geq 0}{j+k\leq \deg m -2}}(j-k)\mu_{j+k+2}\,
p_{ji}\,q_{ki}.
\label{Riiderm}\eq
The identity (\ref{GG}) also requires modification if we do not have $m(x)=1$,
but since we shall only use it in this special case there is no need to write
down the modification.
\subsection{The Exponential Variant}
\label{subsec:DE_circle}
Here we consider kernels of the form
\bq K(x,y)\chJ(y):= {\varphi(x)\psi(y)-\psi(x)\varphi(y)\over e^{bx} - e^{by}}
\chJ(y)\  \label{expker}\eq
where $b$ can be an arbitrary complex number.
Because of the different denominator it turns out that the differentiation
formulas should now be of the form
\begin{eqnarray}
m(x)\varphi'(x)&=& \left(A(x)+\fr{b}{2} m(x)\right) \varphi(x)
+B(x) \psi(x), \nonumber \\
m(x) \psi'(x)&=&-C(x)\varphi(x) -\left(A(x)-\fr{b}{2} m(x)\right)\psi(x)
\label{diff_formulas_circle} \end{eqnarray}
where $A(x)$, $B(x)$, $C(x)$ and $m(x)$ are ``exponential polynomials,''
finite linear combinations of the exponentials $e^{kbx}\;(k=0,\pm 1,\pm 2,
\cdots)$.
We compute, as the analogue of (\ref{KM_deriv}), that
\begin{eqnarray*}
\left( m(x)\fr{\pl}{\pl x}+m(y)\fr{\pl}{\pl y} + \fr{b}{2} \>
\fr{(m(x)-m(y))(e^{bx}+e^{by})}{e^{bx}-e^{by}}\right) K(x,y)
&&   \\
\hspace*{10mm}
=  \mbox{ the right hand side of (\ref{K_deriv}) with
denominator $x-y$}&& \end{eqnarray*}
\vspace*{-12mm}
\bq
\hspace*{-1.25in}\mbox{
 replaced by $e^{bx}-e^{by}$.}  \label{KM_deriv_circle}
\eq
Now, of course, we write
\[ A(x)=\sum_j \al_j e^{bjx}, \ \ \mbox{etc.,} \]
the summations summing over negative and nonnegative indices, and
\bq
\fr{A(x)-A(y)}{e^{bx}-e^{by}}=\sum_{j,k\geq 0} \al_{j+k+1} e^{jbx}
e^{kby} - \sum_{j,k\leq -1} \al_{j+k+1} e^{jbx} e^{kby},\ \mbox{etc.}
\label{A_eqn_circle}
\eq
What arise now are functions $Q_k$,\ldots, $w_k$ defined, for negative
as well as nonnegative values of $k$, by replacing $x^k$ by
$e^{kbx}$ in the earlier definitions. Analogues of (\ref{Qk_recursion})
and (\ref{Pk_recursion}) hold for negative as well as positive values
of $k$ so all the $Q_k$ and $P_k$ are expressible in terms of
$Q_0$, $P_0$, as well as the $\tilde v_k$ in terms of the $u_k$,
$v_k$, $w_k$.
\par
Notice that if $m(x)$ is constant then the third term in the large
parentheses in (\ref{KM_deriv_circle}) vanishes and so we obtain
in the end the analogue of (\ref{qi_ai}) and (\ref{pi_ai}); in addition
to the change in the range of indices now and the fact that the
double sums have two parts, as in (\ref{A_eqn_circle}), we must in
the single sums over $j$ add $\fr{b}{2}\mu_j$ to the terms
$\al_j$ and $\be_j$ in (\ref{qi_ai}) and the terms $-\ga_j$ and
$-\al_j$ in (\ref{pi_ai}). For general $m(x)$ we must insert factors
$m(a_i)$ on the left sides  of (\ref{qi_ai}) and (\ref{pi_ai})
and factors $m(a_k)$ in the last summands on the right, and then add
terms coming from the difference
\bq \fr{1}{e^{bx}-e^{by}}\left[ m'(y) - \fr{b}{2}\> \fr{(m(x)-m(y))(
e^{bx}+e^{by})}{e^{bx}-e^{by}}\right]\left[\>\varphi(x)\psi(y)
- \psi(x)\varphi(y)\>\right] \label{circm}\eq
as at the end of the preceding section.  We shall not write these down
since in the only case we consider later we have $m(x)=1$.
Two of the equations involving $R(x,y)$ must also be modified. We see first
that in (\ref{Rjk}) the denominator must be replaced by
$e^{ba_{j}}-e^{ba_{k}}$.
Second, (\ref{Rxx}) must have the factor $b\,e^{bx}$ inserted on the left side,
with the result that (\ref{Rii}) must have the factor $b\,e^{ba_{i}}$ inserted
on the left side. Note that (\ref{GG}) is unchanged.
\medskip\par\noindent
{\it Remark 1\/}. The product of the first two factors in (\ref{circm}) is an
exponential polynomial in $e^{bx}$ and $e^{by}$. It was precisely to achieve
this outcome that we required the formulas (\ref{diff_formulas_circle}) to
have the form they do.
\medskip\par\noindent
{\it Remark 2\/}. In case $b$ is real the change of variable $x\mapsto e^{bx}$
transforms the operator with kernel (\ref{kernel}) acting on the set $e^{bJ}$
to an operator with kernel of the form (\ref{expker}) acting on $J$, with the
new $(\vph,\psi)$ pair satisfying (\ref{diff_formulas_circle}). So we see
that there is more than simply an anology between the two situations. In fact
we
could have allowed the various coefficients in (\ref{diff_formulas})
to be linear combinations of negative or nonnegative integral powers of $x$,
and
then the two situations would have been completely equivalent for real $b$.

\section{Sine, Airy and Bessel}
\label{sec:sine_airy_bessel}
\subsection{Sine Kernel}
\label{subsec:sine_kernel}
The simplest example is the sine kernel
\[ K(x,y)=\fr{\la}{\pi} \; \fr{\sin(x-y)}{x-y} \]
where we take
\[ \varphi(x)=\st{\fr{\la}{\pi}}\, \sin x, \ \ \
\psi(x)=\st{\fr{\la}{\pi}}\, \cos x. \]
The differentiation formulas hold with
\[ m(x)=1, \ \ \   A(x)=0, \ \ \   B(x)=1, \ \ \  C(x)=1.\]
(It is useful to incorporate a parameter  $\la\in [0,1]$ into $K$; c.f.\
formula (\ref{detform}).)\ \
The  partial differential equations are (\ref{qjk}), (\ref{pjk})
(the universal equations along with universal relation (\ref{Rjk})), and the
specialization  of
(\ref{qi_ai}) and (\ref{pi_ai}) which now read, respectively,
\begin{eqnarray}
\fr{\pl q_i}{\pl a_i}&=&p_i - \sum_{k\neq i} (-1)^k R(a_i,a_k) q_k\, ,
\label{jmms1}  \\
\fr{\pl p_i}{\pl a_i}&=&-q_i - \sum_{k\neq i} (-1)^k R(a_i,a_k) p_k\, ,
\label{jmms2}\end{eqnarray}
along with the specialization of (\ref{Rii})
\bq R(a_i,a_i) = p_i^2+q_i^2 + \sum_{k\neq i} (-1)^k \fr{(q_i p_k - p_i q_k)^2}
{a_i-a_k}\; . \label{jmms3} \eq
These are the  equations of JMMS \cite{jmms} though they appear here in
a slightly  different form due to our use of sines and cosines in the
definitions of $\varphi$ and $\psi$ rather than the alternative choice
of $e^{\pm ix}$, which we could have taken just as well.
(They also appear slightly different  in \cite{tw1}
due to our convention here not to put in a factor of $\pi$ into
the definition of the sine kernel.)\ \
\par
For the case of a single interval $J=(-t,t)$, $s=2t$, these equations
imply that $\sigma(s;\la):=-s R(t,t)$ satisfies the Jimbo-Miwa-Okamoto
$\sigma$ form of Painlev{\'e} V.  We refer the reader to the
literature for a derivation of this,   properties of the solution
of this equation,  and the implications for random matrices
\cite{btw,dyson92,jmms,mehta92a,mehta_mahoux92b,tw1,widom92}.
\subsection{Airy Kernel}
\label{subsec:airy_kernel}
For the Airy kernel we have (again inserting a parameter $\la$ into
$K$)
\[ \varphi(x)=\st{\la}\,  {\rm Ai}(x),
 \ \ \ \psi(x)=\st{\la}\, {\rm Ai}'(x) \]
from which it follows that
\[ m(x)=1, \ \ \ A(x)=0, \ \ \ B(x)=1, \ \ \ C(x)=-x \]
since ${\rm Ai}''(x)=x {\rm Ai}(x)$.
For notational convenience we write $u=u_0$ and $v=v_0$.
In addition to the universal relations (\ref{Rjk})--(\ref{pjk}), we
have two additional equations for $u$ and $v$, viz.\
(\ref{uj_eqn}) and (\ref{vj_eqn}) for $j=0$.
Using the recursion relation (\ref{Qk_recursion}) for $k=1$, we
deduce that (\ref{qi_ai}) and (\ref{pi_ai}) reduce to
\begin{eqnarray}
\fr{\pl q_i}{\pl a_i}&=&-u q_i+p_i-\sum_{k\neq i} (-1)^k R(a_i,a_k) q_k\, ,
\label{airy1} \\
\fr{\pl p_i}{\pl a_i}&=&(a_i-2v)q_i+u p_i -\sum_{k\neq i} (-1)^k R(a_i,a_k)
p_k \, ,
\label{airy2}\end{eqnarray}
and (\ref{Rii}) reduces to (again using the recursion relation
(\ref{Qk_recursion}) for $k=1$)
\bq R(a_i,a_i)=p_i^2-a_i q_i^2-2u q_i p_i +2 v q_i^2 +
\sum_{k\neq i} (-1)^k \fr{(q_i p_k - p_i q_k)^2}
{a_i-a_k}\; . \label{airy3} \eq
These are the equations derived in \cite{tw2}. We mention that
in addition to these equations, two first integrals were derived
which can be used to represent $u$ and $v$ directly in terms
of the $q_j$ and $p_j$ (see (2.18) and (2.19) in \cite{tw2}).
We also remark that in the case $J=(s,\iy)$, the quantity $R(s,\iy)$
was shown to satisfy the second order nonlinear $\sigma$ DE associated
to Painlev{\'e} II.  Again we refer the reader to \cite{tw2} for
details.
\subsection{Bessel Kernel}
\label{subsec:bessel_kernel}
For the Bessel kernel
\[ \varphi(x)=\st{\la}\, J_\al(\st{x}),\ \ \
\psi(x) = x \varphi'(x), \]
from which it follows (using Bessel's equation) that
\[ m(x)=x, \ \ \ A(x)=0,\ \ \ B(x)=1, \ \ \ C(x)=\fr{1}{4}\left(x-\al^2\right).
\]
Again using the recursion relation (\ref{Qk_recursion}), we deduce that
(\ref{qi_ai}) and (\ref{pi_ai}) become,  with the additional insertions
(\ref{m_insertions1}),
\begin{eqnarray*}
a_i \fr{\pl q_i}{\pl a_i}&=& \fr{1}{4} u q_i + p_i -
\sum_{k\neq i} (-1)^k a_k R(a_i,a_k) q_k \, , \\
a_i \fr{\pl p_i}{\pl a_i}&=& \fr{1}{4}\left(\al^2-a_i+2v\right) q_i
-\fr{1}{4} u p_i - \sum_{k\neq i} (-1)^k a_k R(a_i,a_k) p_k\, ,
\end{eqnarray*}
and (\ref{Rii}) with insertions (\ref{mR_insertions1}) becomes
\[ a_i R(a_i,a_i) = -\fr{1}{4}\left(\al^2-a_i+2v\right) q_i
+\fr{1}{2} u q_i p_i + p_i^2 +\sum_{k\neq i} (-1)^k a_k
\fr{(q_ip_k-p_iq_k)^2}{a_i-a_k}\; .
\]
(As before, $u=u_0$ and $v=v_0$.)\ \
These are the equations derived in \cite{tw3}.  As was the case
for the Airy kernel, two first integrals were derived which can
be used to express $u$ and $v$ directly in terms of the $q_j$ and
$p_j$.  For the case $J=(0,s)$, the quantity $\sigma(s)=s R(0,s)$
was shown \cite{tw3} to satisfy the $\sigma$ DE for Painlev{\'e} III
\cite{jm,okamoto87a}.
%%%%%%%%%%%%%%%%%%%%%%%%%%%%%%%%%%%%%%%%%%%%%%%%%%%%%%%%%%%%%%%%%%%%%%%%%%%%%%
\section{Beyond Airy}
\label{sec:beyond_airy}
In this section we give as an example of our general system of partial
differential equations the simplest case ``beyond Airy'' in the sense
discussed in the Introduction.  In the language of 2D quantum gravity
matrix models (see \cite{brezin_review} for a review), we are considering
the case of pure gravity.  Thus we take
\begin{eqnarray*}
 {\cal Q}& =& D_x^2 +\xi(x)\, , \\
 {\cal P}& =&\left( {\cal Q}^{3/2}\right)_+ \\
&=& D_x^3 +\fr{3}{2} \xi(x) D_x + \fr{3}{4} \xi'(x)
\end{eqnarray*}
where $D_x$ is differentiation with respect to $x$.
The string equation implies that
$\xi(x)$  satisfies
\[ \xi'''(x) + 6 \xi(x) \xi'(x) + 4 = 0 \]
which when integrated is  Painlev{\'e} I
\cite{brezin_kazakov,douglas_shenker,gross_migdal}:
\[ \xi''(x)+ 3 \xi^2(x) + 4 x = 0.  \]
(Without loss of generality we may set the constant of integration
to zero since it corresponds simply to a shift in the variable $x$.
And, of course, the `3' and `4' can be changed  by scale transformations
 to give
the canonical form of Painlev{\'e} I.)\ \
Exactly what solution $\xi(x)$ one chooses for pure gravity  is
still of some debate (on this point see \cite{david} and references therein).
The function $\varphi(\la,x)$ satisfies (\ref{schrodinger}) and
(\ref{P_operator}) which implies that if we define $\psi(\la,x)$ by
(\ref{psi_beyond_airy}), then the differentiation formulas are
\begin{eqnarray*}
 m(\la)=1, & A(\la)= -\fr{1}{4}\,  \xi'(x), & B(\la) = \la + \fr{1}{2}\,
 \xi(x), \\
& C(\la) = - \la^2 +\fr{1}{2} \xi(x) \la + \fr{1}{2}\,  \xi^2(x) +\fr{1}{4}\,
 \xi''(x),
&
\end{eqnarray*}
where we remind the reader of the change of notation in the independent
variable (see Introduction).
\par
Since $C(\la)$ is quadratic in $\la$, the equations will involve
$u_j$, $v_j$, and $w_j$ for $j=0,1$.  Using the recusion relations
(\ref{Qk_recursion}), (\ref{Pk_recursion}), (\ref{vktilde_recursion})
the equations (\ref{qi_ai}) and (\ref{pi_ai}) specialize to
\[
\fr{\pl}{\pl a_i}
\left(\begin{array}{c} q_i \\ p_i  \end{array}\right)
 = {\cal M}(a_i)
\left(\begin{array}{c} q_i \\ p_i  \end{array}\right)
-\sum_{k\neq i} (-1)^k R(a_i,a_k)
\left(\begin{array}{c} q_k \\ p_k  \end{array}\right)\]
where ${\cal M}(a_i)$ is the $2\times 2$ matrix whose elements are
given by
\begin{eqnarray*}
{\cal M}_{11}(a_i)& =& -\fr{1}{4}\, \xi'-w  +\fr{1}{2}\, \xi u +u v - u_1 -
a_i u, \\
{\cal M}_{12}(a_i)&=& a_i - u^2 + 2 v + \fr{1}{2} \xi, \\
{\cal M}_{21}(a_i)&=& a_i^2 - \fr{1}{2} \xi^2 - \fr{1}{4} \, \xi'' -2 a_i v
+ 3 v^2 -2 v_1 - 2 u w -\fr{1}{2}\, a_i \xi + v \xi, \\
{\cal M}_{22}(a_i)&=&-{\cal M}_{11}(a_i).
\end{eqnarray*}
Similarly, (\ref{Rii}) specializes to
\[ R(a_i,a_i) =2{\cal M}_{11}(a_i) q_i p_i + {\cal M}_{12}(a_i)  p_i^2
- {\cal M}_{21}(a_i) q_i^2
+\sum_{k\neq i} (-1)^k \fr{(q_i p_k - p_i q_k)^2}
{a_i-a_k}\; . \]
The universal equations are (\ref{Rjk})--(\ref{pjk}) and
 (\ref{uj_eqn})--(\ref{wj_eqn}).
\par
For the  case $J=(s,\iy)$ (this should be compared with the
analagous case in Airy \cite{tw2}), we are able to find two first
integrals that allow us to eliminate the quantities $u_1$ and $v_1$.
(It is natural to take the boundary condition that all quantities
evaluated at $\iy$ vanish.)\ \   We  denote by $q=q(s)$, etc.\
the quantities corresponding to  the first endpoint $a_1=s$.
The first relation is quite simple
\[ q^2 - 2 u u_1 + v^2 + 2 v_1 +\fr{1}{2} \xi u^2 + \xi v -
\fr{1}{2} \xi' u = 0, \]
but the second one we found is rather messy
\begin{eqnarray*}
-p^2 + s q^2 + u + 2 p q u + x u^2 + q^2 u^2 - 2 u^3 u_1 - u_1^2
- 2 x v \\
 - 4 q^2 v + 6 u u_1 v
 + 3 u^2 v^2 - 8 v^3 - 2 u^3 w - 2 u_1 w + 6 u v w - w^2 \\
+ \xi\left( -q^2
+\fr{1}{2}  u^4 +2 u u_1 - 6 v^2 + 2 u w - \fr{1}{4} \xi  u^2
- \fr{3}{2} \xi v \right) \\
+ \xi' \left( -\fr{1}{2} u^3 - \fr{1}{2} u_1 +\fr{3}{2} u v
- \fr{1}{2} w + \fr{1}{2} \xi u \right) = 0
\end{eqnarray*}
where $q$, $p$, $u$, etc.\ have argument $s$
and $\xi$ has argument $x$.  (The
variable  $x$ appears since
we used Painlev{\'e}~I to eliminate $\xi''(x)$ in our equations.)
\par
Using these first integrals
to eliminate $u_1$ and $v_1$
(note that $w$ also drops out)  we obtain the system of equations:
\begin{eqnarray*}
q''&=&\left(x s + s^3 + \fr{1}{2} x \xi +\fr{1}{8} \xi^3
+\fr{1}{16} (\xi')^2\right)q + p+  \left(4 s - 4 v - \xi\right) q^3
+4 u  q^2 p - 2 q p^2, \\
p''&=& \left(2s-2v -\fr{1}{2} \xi \right) q+
\left(x s + s^3 + \fr{1}{2} x \xi+\fr{1}{8} \xi^3 +\fr{1}{16} (\xi')^2
+ 2 u\right)p \\
&& + \left(4 s- 4v- \xi\right) p q^2 +4 u q p^2 - 2 p^3
\end{eqnarray*}
and, of course, we still have the universal equations
\[ u'=-q^2, \ \ \ v'=-q p. \]
Letting  $R(s)=R(s,\iy)$ we find from (\ref{Riidform}) and (\ref{Riiderm}) that
\[ R'= \left( - 2 s+2 v+ \fr{1}{2} \xi\right)  q^2 + p^2 - 2 u q p . \]
Clearly, further  analysis of these equations is needed to be able to
analyze the associated Fredholm determinant.
For example, can one derive a differential equation for $R$ itself?

\section{Finite $N$ Hermite, Laguerre, Jacobi  and Circular}
\label{sec:her_lag_jac}
\subsection{Hermite Kernel}
\label{subsec:hermite_kernel}
\subsubsection{The partial differential equations and a first integral}
\label{subsubsec:hermite_pde}
It follows from the Christoffel-Darboux formula that the kernel
$\la\,K_{N}(x,y)$ for the  finite $N$ Gaussian Unitary Ensemble (GUE)
is of the form (\ref{kernel}) provided we choose
\[ \varphi(x)=\la^{1/2} \left(\fr{N}{2}\right)^{1/4} \varphi_N(x), \ \ \
\psi(x)=\la^{1/2} \left(\fr{N}{2}\right)^{1/4} \varphi_{N-1}(x) \]
with  $\varphi_k(x)$  the harmonic oscillator wave functions
\[ \varphi_k(x) = \fr{1}{\st{2^k k! \st{\pi}}}
\, e^{-x^2/2} H_k(x), \ \ \ k=0,1,\ldots \]
and  $H_k(x)$ are the Hermite polynomials \cite{erdelyi}. This is
well-known and we refer the reader to \cite{mehta_book} for details.
{}From the differentiation and
recurrence formulas for Hermite polynomials it follows that the
differentiation formulas for $\varphi$ and $\psi$  hold with
\[ m(x)=1, \ \ \ A(x)=-x, \ \ \ B(x)=C(x)=\st{2N}. \]
We can therefore immediately write down the equations
\begin{eqnarray}
\frac{\pl q_{j}}{\pl a_{j}}&=&-a_j q_j + \left( \st{2N}-2u\right)p_j
-\,\sum_{k\neq j}(-1)^{k}\,\frac{q_{j}p_{k}-p_{j}q_{k}}
{a_{j}-a_{k}}\,q_{k}\, ,
\label{hermite_qj} \\ \noalign{\vskip3pt}
\frac{\pl p_{j}}{\pl a_{j}}&=&a_j p_j -  \left( \st{2N}+2w\right) q_j
-\sum_{k\neq j}(-1)^{k}\,
\frac{q_{j}p_{k}-p_{j}q_{k}}
{a_{j}-a_{k}}\,p_{k} \label{hermite_pj}
\end{eqnarray}
along with
\bq
R(a_j,a_j)=-2 a_j p_j q_j + \left( \st{2N}-2 u\right) p_j^2 +
  \left( \st{2N}+2w\right) q_j^2
+\sum_{k\neq j} (-1)^k{(q_j p_k - p_j q_k)^2 \over a_j - a_k } \, ,
\label{hermite_Rjj}\eq
\bq \fr{\pl}{\pl a_{j}}R(a_{j},a_{j})=-2\,p_{j}\,q_{j}
-\sum_{k\neq j}(-1)^{k}R(a_{j},a_{k})^{2}\, ,\label{hermiteRjjder} \eq
\bq \fr{d}{dt}R_{t\,J}(t\,a_{j},t\,a_{j})=-2\,a_{j}\,p_{j}\,q_{j}
+\sum_{k\neq j}(-1)^{k}(a_{k}-a_{j})R_{t\,J}(t a_{j},t a_{k})^{2}\, .
\label{hermiteGG} \eq
These follow from formulas (\ref{qi_ai})--(\ref{GG}).
\par
We now  derive a first integral involving $u$, $w$, $p_j$ and
$q_j$; namely we show
\bq
\st{2N}\left(u-w\right) + 2 u w = - \sum_{j} (-1)^j p_j q_j .
\label{hermite_1stint}\eq
Observe that
\[
- \left(\sum_k\fr{\pl}{\pl a_k}\right) p_j q_j =
\left(\st{2N} + 2w\right) q_j^2 - \left(\st{2N}-2u\right) p_j^2 \]
and
\[ \fr{\pl}{\pl a_j}\left( \st{2N}\left(u-w\right) + 2 u w\right)
=(-1)^j \left(\st{2N} + 2w\right) q_j^2 -
 (-1)^j\left(\st{2N}-2u\right) p_j^2. \]
Multiplying the first equation
 by $(-1)^j$ and summing both equations over $j$,
 results in
\[
\left(\sum_k \fr{\pl}{\pl a_k}\right)\left(-\sum_j (-1)^j p_j q_j \right) =
\left(\sum_k \fr{\pl}{\pl a_k}\right) \left(\st{2N}\left(u-w\right) + 2 u w
\right). \]
It follows that the two sides of (\ref{hermite_1stint}) differ by a function
of $(a_1,\ldots,a_{2m})$ which is invariant under translation by any
vector $(s,\cdots,s)$.  Since, clearly, both sides tend to zero as all
$a_i\ra\iy$, their difference must be identically zero.
\subsubsection{Bulk scaling limit of finite N equations}
We now show how (\ref{hermite_qj})--(\ref{hermite_Rjj}) reduce
to the sine kernel equations (\ref{jmms1})--(\ref{jmms3}) in the
``bulk scaling limit.''  For a fixed point $z$, i.e.\
independent of $N$, the density $\rho(z)$  in the GUE is asymptotic to
 $\st{2N}/\pi $ as $N\ra \iy$.  The bulk scaling limit
corresponds to measuring fluctuations about this fixed point $z$ on
a stretched length scale proportional to $\st{2N}$
and then taking  $N\ra\iy$.  Denoting for
the moment the bulk quantities with a superscript $B$, this means
we set
\[ a_j = z + \fr{a_j^B}{\st{2N}} \]
and consider the limit  $N\ra\iy$, $a_j\ra z$ such that $a_j^B$ is
fixed and ${\rm O}(1)$.
In this limit we deduce from the asymptotics
of the harmonic oscillator wave functions (see, e.g., Appendix 10
in \cite{mehta_book}) that both $\varphi$ and $\psi$ are ${\rm O}(1)$
quantities in the bulk scaling limit.  From this and the
fact that it is $K(x,y) dy$ which is ${\rm O}(1)$, we deduce that
both $q_j$ and $p_j$ are ${\rm O}(1)$ quantities in the bulk scaling
limit.  An examination of the inner products defining both $u$ and $v$
shows that these too are ${\rm O}(1)$ quantities.
Thus if we formally replace
\[ a_j \ra  z + \fr{a_j^B}{\st{2N}},\ \ \  q_j\ra q_j^B, \ \ \
p_j\ra p_j^B, \ \ \ R(a_j,a_j)\ra \st{2N}\, R^B(a_j^B,a_j^B) \]
in (\ref{hermite_qj})--(\ref{hermite_Rjj}) (and replace all derivatives
by derivatives with respect to $a_j^B$), take $N\ra\iy$,  we obtain
(\ref{jmms1})--(\ref{jmms3}).
\subsubsection{Semi-infinite interval and Painlev{\'e} IV}
In this section we specialize
the finite $N$ GUE equations  to the  case of $m=1$,  $a_1=s$ and
$a_2=\iy$, i.e.\ $J=(s,\iy)$. We write $q(s)$, $p(s)$, and $R(s)$ for
$q_1$, $p_1$, and $R(a_1,a_1)$,
respectively,  of the previous section.  The
differential  equations reduce to
($^\prime = d/ds$)
\begin{eqnarray}
q'&=& -s q +\left(\st{2N} - 2u\right) p, \label{hermite_dq/ds} \\
p'&=& s p -\left(\st{2N} + 2w\right) q, \label{hermite_dp/ds} \\
u'&=& - q^2, \label{hermite_du/ds} \\
w'&=& - p^2, \label{hermite_dw/ds} \end{eqnarray}
(\ref{hermite_Rjj}) reduces to
\bq R(s) = -2 s p q +\left(\st{2N} - 2u\right) p^2 +
\left(\st{2N} + 2w\right) q^2, \label{hermite_R(s)}\eq
and the first integral is now
\bq \st{2N}\left(u-w\right) + 2 u w = p q. \label{hermite_1stint_s} \eq
\par
We proceed to derive a second order
differential equation for $R(s)$ and show that
it is a special case of the Jimbo-Miwa-Okamoto $\sigma$ form
of Painlev{\'e} IV \cite{jm,okamoto86}.
Relation (\ref{hermiteRjjder}) is now
\bq  R'= - 2 p q,\label{hermite_R'(s)} \eq
while (\ref{hermite_dq/ds}) and (\ref{hermite_dp/ds}) give
\[ (pq)' =  \left(\st{2N} - 2u\right) p^2 -
\left(\st{2N} + 2w\right) q^2\, . \]
Differentiating one more time gives
\begin{eqnarray*}
(pq)''&=&2s \left(\st{2N} - 2u\right) p^2 +
\left(\st{2N} + 2w\right) q^2 - 8pq\left( N + \st{2N}\, (w-u) - 2uw\right)
+4 p^2 q ^2  \\
&=& 2s\left\lbrace \left(\st{2N} - 2u\right) p^2 +
\left(\st{2N} + 2w\right) q^2\right\rbrace - 8 N p q + 12 p^2 q^2
\end{eqnarray*}
where we used the first integral (\ref{hermite_1stint_s}) to obtain the second
equality.  Referring back to (\ref{hermite_R(s)}) we see that the term
in curly brackets in the last expression
is  $R + 2spq$.
Using (\ref{hermite_R'(s)}) to eliminate all terms involving $pq$ in the
last equation we find
\[ R''' = -4s(R-sR') - 8N R' - 6 (R')^2. \]
This third order equation can be integrated (the constant of integration
is zero) to give
\bq
(R'')^2+4(R')^2\left(R'+2N\right)-4\left(sR'-R\right)^2 = 0.
\label{hermite_R_IV}
\eq
Comparing this  with (C.37) of \cite{jm} (see also \cite{okamoto86}), we
see immediately that this is the $\sigma$ version of Painlev{\'e} IV
with parameters (in notation of \cite{jm}) $\nu_1=0$ and
$\nu_2=2N$. Explicitly in terms of the Painlev{\'e} IV transcendent
$y=y(s)$ we have
\bq
R(s) = N y -\fr{s^2}{2} y - \fr{s}{2} y^2 - \fr{1}{8} y^3
+ \fr{1}{8 y} (y')^2
\label{hermite_R_P4} \eq
with $P_{IV}$ parameters $\alpha=2N-1$ and $\beta=0$.  Recall
that $w=w(z)$ is a Painlev{\'e} IV transcendent
with parameters $\al$ and $\beta$  if it satisfies
the $P_{IV}$ equation
\[ \fr{d^2 w}{dz^2}=\fr{1}{2w}\left(\fr{dw}{dz}\right)^2 +\fr{3}{2} w^3
+4 z w^2 + 2(z^2-\al)w + \fr{\beta}{w} \, . \]
We are, of course, interested in the family of solutions that vanish
as $s\ra +\iy$.
\par
This particular $P_{IV}$ has been  studied by
Bassom, et.~al.~\cite{bassom} (see also \cite{clarkson_mcleod}). To make
contact with their notation
define
\[ \eta(\xi):= 2^{-3/4}\st{y(s)}, \ \xi:=\st{2}\,  s \]
($y$ is the above $P_{IV}$ transcendent) so that $\eta$ satisfies
\[\fr{d^2\eta}{d\xi^2}=3\eta^5 +2\xi\eta^3+\left(\fr{\xi^2}{4}-\nu-
\fr{1}{2}\right)\eta. \]
with $\nu=N-1$.
They analyze the one-parameter family of solutions $\eta_k(\xi;\nu)$
satisfying the boundary condition $\eta(\iy)=0$.  The parameter $k$
is defined uniquely by the asymptotic condition:
\[ \eta_k(\xi;\nu) \sim k\, \xi^\nu \exp\left(-\fr{\xi^2}{4}\right)
\ \ \ {\rm as} \ \ \ \xi\ra\iy. \]
In terms of our  parameter $\lambda$  we have
\[  k^2=\fr{\la}{2^{3/2}(N-1)!\st{\pi}}\, . \]
(This   identity is derived by examining the large
positive  $s$ asymptotics
of $R(s)$, which is easy because of the rapid decrease
of the kernel as $s\ra +\iy$.)\ \
These authors prove that for all positive integers $N>1$ the solution
$\eta_k(\xi,N-1)$ exists for all $\xi$ whenever $\lambda<1$, and
that $\eta_k(\xi,N-1)$ blows up for a finite $\xi$ whenever $\lambda>1$.
These results are in complete agreement with what one expects from
the spectral theory of the Fredholm determinant.  There are formal,
but not rigorous,  results that solve the connection problem for the
asymptotics as $\xi\ra -\iy$; in particular, for $\la=1$
\bq  \eta_k(\xi;N-1) \sim \left(-\fr{\xi}{2}\right)^{1/2} \ \ \ {\rm as} \ \ \
\xi\ra - \iy . \label{eta_asy} \eq
\par
Using (\ref{eta_asy}) (and computing higher order terms by using the
differential equation) we find that
\bq
 R(s) = -2 N s - \fr{N}{s} + \fr{N^2}{s^3} - \fr{N^2(1+9N^2)}{4 s^5}+
\fr{N^3(10+27N^2)}{4 s^7} + \cdots \ \ \ {\rm as} \ \ \ s\ra-\iy.
\label{hermite_R(s)_asy} \eq
\subsubsection{Distribution functions for $\la_{max}$ and $\la_{min}$}
If we denote the smallest and largest eigenvalues of a matrix from the GUE by
$\la_{\mbox{\footnotesize min}}$ and $\la_{\mbox{\footnotesize max}}$,
respectively, then in the notation of (\ref{detform}) we have
\[ P(\la_{\mbox{\footnotesize max}}<s)=E(0;(s, \iy))=\mbox{det}\,(I-K). \]
Thus using (\ref{logdet}) we deduce the representation
\[ P(\la_{\mbox{\footnotesize
max}}<s)=\exp\left\{-\int_{s}^{\iy}R(t;1)\,dt\right\} \]
where $R(s;\la)$ denotes the function $R(s)$ of the preceding section
with parameter value $\la$. This
is our representation of the distribution function for
$\la_{\mbox{\footnotesize max}}$ in terms of a Painlev\'{e}
transcendent. There is of course a similar representation for the distribution
function for $\la_{\mbox{\footnotesize min}}$.
\par
The authors of \cite{bassom,clarkson_mcleod} give an algorithm to compute
the quantities $\eta_k(\xi,\nu)$, $\nu=$~positive integer, of the last section
 exactly in terms of the error function
\[ I(\xi)=2\int_\xi^\iy \exp\left(-\fr{x^2}{2}\right)\, dx\, . \]
That such elementary solutions of the $P_{IV}$ transcendent exist,
at least for the case $\lambda=1$, is now clear from the random matrices
point of view since  $E(0;(s,\iy))$ is expressible in terms of integrals of
the form
\[ \int_{-\iy}^s x^j e^{-x^2} \, dx\, . \]
This follows from (\ref{expected}) with $f$ the characteristic function of
$(-\iy, s)$.
\subsubsection{Edge scaling limit from Painlev{\'e} IV equation}
The edge scaling limit \cite{tw2} corresponds to the
replacements
\[ s\ra \st{2N}+ \fr{s}{\st{2} N^{1/6}} \ \ \ {\rm and} \ \ \
R\ra \st{2} N^{1/6} R  \]
in  (\ref{hermite_R_IV}) and retaining only the leading order
term as $N\ra \iy$.  The result of doing this is
\bq \left(R''\right)^2 + 4\left(R'\right)^3 + 4 R'\left(R-sR'\right)=0
\label{hermite_R_II}\eq
which is the equation derived in \cite{tw2}.
We remark
that (\ref{hermite_R_II})
 is the $\sigma$ form for Painlev{\'e} II, see (C.17) in \cite{jm}
and Proposition 1.1 in \cite{okamoto86}.
%%%%%%%%%%%%%%%%%%%%%%%%%%%%%%%%%%%%%%%%%%%%%%%%%%%%%%%%%%%%%%%%%%%%%%%%%%%
\subsubsection{Symmetric single interval  case}
In this section we specialize
the finite $N$ GUE equations  to the case of $m=1$, $a_1=-t$ and
$a_2=t$, i.e.\ $J=(-t,t)$. We denote by $q(t)$ and $p(t)$ the quantities
$q_2$ and $p_2$, respectively.  Since
$\varphi_N(-x)=(-1)^N\varphi_N(x)$ and $K(-x,-y)=K(x,y)$,
we have $q_1=(-1)^N q$ and $p_1=-(-1)^N p$.
We further set
\[ R(t):=R(t,t)=R(-t,-t),\qquad \tilde R(t):=(-1)^{N}R(-t,t)=(-1)^{N}R(t,-t)\]
and record that
\[\fr{d \log D(J;\la)}{dt} = - 2 R(t). \]
Now $\varphi$ is even or odd depending on whether $N$ is even or odd, with
$\psi$ having the opposite parity. It follows from this fact, and our choice
of sign in the definition of $\tilde R$, that (\ref{Rjk}) specializes in
this case to
\bq \tilde R(t)=-\fr{q\,p}{t} \label{tildeRrep} \eq
while (\ref{hermite_Rjj}) and (\ref{hermiteGG}) specialize to
\bq R(t) =  -2 t p q +\left(\st{2N}-2u\right) p^2 +\left(\st{2N}+2w\right) q^2
-\fr{2}{t} p^2 q^2 \label{hermite_R(t)}\, , \eq
\bq \fr{dR}{dt} = 2 t \tilde  R + 2 \tilde R^2\, .\label{hermite_gaudin} \eq
The last is the finite $N$ analogue of the Gaudin relation.
(See, e.g.~\cite{mehta_book,tw1}.)
The differential equations  specialize to
\begin{eqnarray}
\fr{dq}{dt}&=&-\fr{\pl q_2}{\pl a_1}+\fr{\pl q_2}{\pl a_2}=
 2 \tilde R \, q - t\,  q +\left(\st{2N}-2u\right) p, \label{hermite_dq/dt} \\
\fr{dp}{dt}&=& -2 \tilde R\,  p +
 t\,  p - \left(\st{2N}+2w\right) q, \label{hermite_dp/dt} \\
\fr{du}{dt}&=& 2 q^2, \label{hermite_du/dt} \\
\fr{dw}{dt}&=& 2 p^2. \label{hermite_dw/dt} \end{eqnarray}
And the first integral (\ref{hermite_1stint}) is now
\bq
\st{2N} (u-w) + 2 u w = - 2 p q. \label{hermite_1stint_t} \eq
\subsubsection{Differential equations for $R$ and $\tilde R$}
It follows easily from (\ref{tildeRrep}), (\ref{hermite_R(t)}) and
(\ref{hermite_gaudin}) that
\begin{eqnarray}
{d\over dt}(t \tilde R) &=& \Re (r^2), \label{hermite_mehta1} \\
{d\over dt}(t R ) &=& \vert r \vert^2 + 4 t^2 \tilde R \label{hermite_mehta2}
\end{eqnarray}
where
\[ r=q\st{\st{2N}+2w}+ip\st{\st{2N}-2u} .\]
Equations (\ref{hermite_mehta1}) and (\ref{hermite_mehta2}) are the finite $N$
analogue of those derived by Mehta \cite{mehta92a}. (See also
discussion in \cite{tw1}.)
\par
We now eliminate the quantity $r$.  For this derivation only, we
write   $a(t):=t R(t)$ and $b(t):=t \tilde R(t)$.  We begin with
the obvious
\bq  \vert r \vert^4 = \Re(r^2)^2+\Im(r^2)^2.\label{r_identity}\eq
Now
\begin{eqnarray}
 \Im(r^2)^2&=& 4 p^2 q^2 \left(\st{2N}-2u\right)\left(\st{2N}+2w\right)
\nonumber \\
&=& 4 p^2 q^2 \left(2N + 4 p q\right)\nonumber \\
&= & 4 b^2 \left(2N-4b\right).\label{Im(r)} \end{eqnarray}
We now use (\ref{hermite_mehta1}) and (\ref{hermite_mehta2})
 to obtain expressions
for $\Re(r^2)^2$ and $\vert r \vert^4$, respectively, and the above
identity for $\Im(r^2)^2$.  These expressions, when used
in (\ref{r_identity}),  we have  an equation for $a$, $b$, and
their first derivatives.
 If we use the generalized Gaudin relation (\ref{hermite_gaudin})
to eliminate the appearance of $da/dt$ (the one appearing to the first
power), we obtain
\bq \left(\fr{da}{dt}\right)^2 = 8a b+8Nb^2 + \left(\fr{db}{dt}\right
)^2 \label{ab_eq1} \eq
and together with (\ref{hermite_gaudin}), which in the $a$ and $b$ variables
reads
\bq
t\left(\fr{da}{dt}\right) = a + 2 b^2 + 2 t^2 b, \label{ab_eq2} \eq
we have two  differential equations for $R$ and $\tilde R$.
\par
Eliminating $a$,
 we obtain a single second order equation
for $b$ and therefore  $\tilde R$:
\bq
\left(t \tilde R'' + 2 \tilde R' - 24 \, t^2\,  \tilde R^2 +
 8 N\, t\, \tilde R
\right)^2 - 4 \left(2\tilde R - t\right)^2\left( 8 t^2 \tilde R^2\left(
N-2 t \tilde R\right)+\left( \tilde R + t \tilde R'\right)^2\right)=0.
\label{tildeR_eq} \eq
This last
equation is the finite $N$ analogue of (1.18) of Mahoux and Mehta
\cite{mehta_mahoux92b}.
We could, in a similar way, eliminate $b$ and so obtain a second order equation
for
$R$, but the result is messy and we shall not write it down.
\subsubsection{Small $t$ expansions of $R$ and $\tilde R$ }
The boundary conditions at $t=0$
 for (\ref{hermite_gaudin}) and (\ref{tildeR_eq}) follow
from an examination of the Neumann expansion of the resolvent kernel.
Setting  $\rho_0:=K(0,0)$, the density of eigenvalues at $0$,  we find
\begin{eqnarray}
R(t)&=& \rho_0 + 2 \rho_0^2 t +\rho_0\left(4\rho_0^2+(-1)^N \right) t^2
+\fr{8\rho_0^2}{9} \left(9\rho_0^2 - 2N+(-1)^N\right) t^3  +
\nonumber \\
&& \ \ \fr{\rho_0}{18} \left( 288 \rho_0^4+40\rho_0^2\left((-1)^N-2N\right)
-12(-1)^N N - 3\right) t^4 + {\rm O}(t^5)
\label{R(t)_small} \end{eqnarray}
and
\begin{eqnarray}
 (-1)^N \tilde R(t)&=& \rho_0 + 2 \rho_0^2 t +
  \fr{\rho_0}{3}\left(12\rho_0^2  -4N - (-1)^N\right) t^2
+\fr{8 \rho_0^2}{9}\left( 9 \rho_0^2-2N+(-1)^N\right) t^3+ \nonumber \\
&&  \fr{\rho_0}{90}\left( 1440 \rho_0^4 +200 \rho_0^2\left((-1)^N-2N\right)
+48N^2+12(-1)^N N +9 \right)t^4 + {\rm O}(t^5).\nonumber \\
&&
\label{tildeR(t)_small} \end{eqnarray}
\subsubsection{Level spacing probability density $p_N(t)$ }
For $m=1$ if we let  $E_N(0;a_1,a_2)$ denote the probability that no
eigenvalues lie in the
interval $(a_1,a_2)$  and $p_N(0;a_1,a_2)\, da_2$ the conditional
probability that given an eigenvalue at $a_1$ the next one lies
between $a_2$ and $a_2+da_2$, then the two quantities are related
by
\[ p_N(0;a_1,a_2)=\fr{1}{\rho(a_1)}{\pl^2 E_N(0;a_1,a_2)\over
\pl a_1 \pl a_2} \]
where $\rho(a_1)$ is the density of eigenvalues at $a_1$.  From
the expression for the logarithmic derivative
of the determinant (with $\la=1$),  we have
\[
{\pl^2 E_N(0;a_1,a_2)\over\pl a_1 \pl a_2}=
{\pl R(a_1,a_1)\over \pl a_2} E_N - R(a_1,a_1) R(a_2,a_2) E_N \, .\]
Differentiating (\ref{hermite_Rjj}) (with $m=1$ and $j=1$) with respect
to $a_2$  we obtain
\[ {\pl R(a_1,a_1)\over \pl a_2}=\left(\fr{q_1 p_2 - p_1 q_2}{a_1-a_2}
\right)^2 \]
which when evaluated at $a_1=-t$, $a_2=t$ becomes
\[{\pl R(a_1,a_1)\over \pl a_2}\biggr\vert_{a_1=-t,a_2=t}
=\left(\tilde R \right)^2 \, .\]
Calling $p_N(t):=p_N(0;-t,t)$ we thus obtain
\[ p_N(t) = \fr{1}{\rho(t)} \left( \tilde R^2(t) - R^2(t) \right) E_N(t) \]
where $E_N(t)=E_N(0;-t,t)$ (we used also $\rho(-t)=\rho(t)$).
\par
Using the expansions
of $R$, $\tilde R$ and $E_N(t)$ we find
\begin{eqnarray*}
 p_N(t)&=&\fr{8}{3}\left(N+(-1)^N\right)\rho_0 t^2
-\fr{8}{45}\left(16N^2+29 (-1)^N N+ 13\right) \rho_0 t^4 + \\
&& \ \
\fr{4}{315}\left(128 N^3 +452 (-1)^N N^2 + 529 N + (-1)^N 205\right) \rho_0
t^6 + \cdots\,. \end{eqnarray*}
Not only does this hold for fixed $N$ and $t$, but it also holds  uniformly
in $N$ and $t$
as long as $t=O(N^{-1/2})$. The reason is that in this range of the
parameters the operator $K$ has norm less than a constant which is less than 1
and has bounded Hilbert-Schmidt norm. Thus the Neumann series for the resolvent
kernel converges in  trace norm.
\par To compare with the bulk scaling limit
 we replace $t$ by $t/\rho_0$,
and deduce (recall $\rho_0\sim \st{2N}/\pi$ as $N\ra \iy$) that
\[ p(t):=\lim_{N\ra \iy} \fr{1}{\rho_0} p_N(\fr{t}{\rho_0})
= \fr{\pi^2}{3} t^2 -\fr{2\pi^4}{45} t^4 +\fr{\pi^6}{315} t^6 + \cdots \]
which is the well-known result \cite{mehta_book}.  Observe that the
large $N$ corrections to these limiting coefficients are ${\rm O}(1/N)$.
(Note that we inserted a factor of $\pi$ in our definition of the
new $t$ variable so as to have the same normalization as in \cite{mehta_book}.)
%%%%%%%%%%%%%%%%%%%%%%%%%%%%%%%%%%%%%%%%%%%%%%%%%%%%%%%%%%%%%%%%%%%%%%%%%%%
\subsection{Laguerre Kernel}
\subsubsection{The partial differential equations}
Again by the Christoffel-Darboux formula it follows that the kernel
for the finite $N$ Laguerre Ensemble of $N\times N$ hermitian matrices
is of the form (\ref{kernel}) provided we choose
\bq \varphi(x)=\st{\la}\, \left(N(N+\al)\right)^{1/4} \varphi_{N-1}(x),
\ \ \ \psi(x)=\st{\la}\, \left(N(N+\al)\right)^{1/4} \varphi_{N}(x)
\label{lagppdef} \eq
($a_{N}$ in (\ref{christoffel}) is negative now) where
\[ \varphi_k(x) =\st{\fr{k!}{\Gamma(k+\al+1)}}\, x^{\al/2} \, e^{-x/2}
L_k^{\al}(x)\, , \]
and $L_k^{\al}(x)$ are the (generalized)  Laguerre polynomials \cite{erdelyi}.
See Chap.~19 of \cite{mehta_book}
and \cite{nagao_wadati}  for further details and references.
{}From the differentiation and recurrence formulas for Laguerre polynomials
it follows that we have differentiation formulas (\ref{diff_formulas}) for
$\varphi$
and $\psi$ with
\[ m(x)=x, \ \ \ A(x)=\fr{1}{2} x -\fr{\al}{2}-N, \ \ \
B(x)=C(x)=\st{N(N+\al)} \, . \]
We therefore have the equations
\begin{eqnarray}
a_j\, \fr{\pl q_j}{\pl a_j}&=&\left(\fr{1}{2} a_j -\fr{\al}{2} - N \right) q_j
+\left(\st{N(N+\al)} + u \right) p_j \nonumber \\
&& - \sum_{k\neq j} (-1)^k a_k R(a_j,a_k) q_k \, , \label{laguerre_qj}\\
a_j \, \fr{\pl p_j}{\pl a_j}&=& - \left(\st{N(N+\al)} - w \right) q_j
-\left(\fr{1}{2} a_j -\fr{\al}{2} - N \right) p_j \nonumber \\
&& - \sum_{k\neq j} (-1)^k a_k R(a_j,a_k) p_k \, , \label{laguerre_pj} \\
a_j\, R(a_j,a_j)&=&\left( a_j - \al - 2N\right) q_j p_j
+\left(\st{N(N+\al)}+u\right) p_j^2 +\left(\st{N(N+\al)}-w\right) q_j^2
\nonumber \\
&& + \sum_{k\neq j} (-1)^k a_k  \fr{(q_j p_k - p_j q_k)^2}{a_j-a_k} \; ,
\label{laguerre_Rjj} \\
\fr{\pl}{\pl a_{j}}a_j\, R(a_j,a_j)&=&q_{j}\,p_{j}-\sum_{k\neq j}
(-1)^{k}a_{k}R(a_{j},a_{k})^{2}. \label{laguerreRjjder}  \end{eqnarray}
These follow from formulas (\ref{qi_ai})--(\ref{Riidform}) as modified by
(\ref{m_insertions1}) and (\ref{mR_insertions1}).
\subsubsection{Single Interval Cases $(0,s)$ and $(s,\iy)$}
We consider first the interval $(0,s)$. We set $a_{1}=0, a_{2}=s, q_{2}=q,
p_{2}=p,
R(s,s)=R(s)$ and
find that eqs. (\ref{laguerre_qj})--(\ref{laguerreRjjder}) with $j=2$
specialize to
\begin{eqnarray}
s\,q'&=&\left(\fr{1}{2} s -\fr{\al}{2} - N \right) q
+\left(\st{N(N+\al)} + u \right) p, \label{laguerre_q0s} \\
s \, p'&=& - \left(\st{N(N+\al)} - w \right) q
-\left(\fr{1}{2} s -\fr{\al}{2} - N \right) p, \label{laguerre_p0s} \\
s\, R(s)&=&\left(s-\al-2N\right) q\, p+\left(\st{N(N+\al)}+u\right) p^2\\
&& +\left(\st{N(N+\al)}-w\right) q^2,\label{laguerre_R0s}\\
(s\,R(s))'&=&q\,p \label{laguerreRder0s}
\end{eqnarray}
(notice that the terms corresponding to $k=1$ in the double sums on the right
sides
of (\ref{laguerre_qj})--(\ref{laguerreRjjder}) are equal to zero),
while (\ref{uj_eqn}) and (\ref{wj_eqn}) specialize to
\bq u'=q^{2},\quad w'=p^{2}. \label{uv0s} \eq
Tedious but straightforward computation using
(\ref{laguerre_q0s})--(\ref{uv0s}) gives
\bq s(pq)'=\left(\st{N(N+\al)} + u \right) p^{2}-\left(\st{N(N+\al)} - w
\right) q^{2},
\label{laguerrepqder} \eq
\begin{eqnarray}
s^{2}(pq)''&=&(2N+\al -s)\left\{\left(\st{N(N+\al)} + u \right)
p^{2}+\left(\st{N(N+\al)}
- w \right) q^{2}\right\} \nonumber \\
&&-\left\{\left(\st{N(N+\al)} + u \right) p^{2}-\left(\st{N(N+\al)}
- w \right) q^{2}\right\}\,+2\,s\,p^{2}\,q^{2} \nonumber \\
&&-4\,N\,(N+\al)\,p\,q+\left\{u\,w+\st{N(N+\al)}\,(w-u)\right\}4\,p\,q.
\label{laguerrepqder2} \end{eqnarray}
Now it follows from (\ref{laguerreRder0s}) and
(\ref{laguerrepqder}) that
\bq s\,R(s)-s\,q\,p \label{sR-spq} \eq
has derivative
\[ \left(\st{N(N+\al)} + u \right) p^{2}-\left(\st{N(N+\al)} - w \right) q^{2}.
\]
But it follows from (\ref{uv0s}) that
\[ u\,w+\st{N(N+\al)}\,(w-u) \]
has exactly the same derivative. Hence the two must differ by a constant.
This constant must be 0 since
(\ref{sR-spq}) clearly vanishes when $s=0$, and so do $u$
and $w$. Thus we have derived the identity
\bq u\,w+\st{N(N+\al)}\,(w-u) = s\,R(s)-s\,q\,p. \label{uwid} \eq

Now we can see that that every term in (\ref{laguerrepqder2}) can be expressed
in terms
of $R(s)$ and its derivatives (up to order 3). By (\ref{laguerreRder0s}) this
is clear
for all products $p\,q$ and its derivatives. This is true of the first
expression in
curly brackets in (\ref{laguerrepqder2}) by what we just said and
(\ref{laguerre_R0s}),
of the first expression in curly brackets by (\ref{laguerrepqder}), and of the
last
expression in curly brackets by (\ref{uwid}).

Thus we have derived a third-order differential equation for $R(s)$. In terms
of
$\si(s):=s\,R(s)$ it reads
\bq s^{2}\si'''=(2\,N+\al-s)\,\si+(\al^{2}+s^{2}-4\,N\,s-2\,\al\,s)\,\si'
-s\,\si''+6\,s\,(\si')^{2}-4\,\si\,\si'.\label{lag3rd} \eq
It follows from this that the two sides of a purported identity
\begin{eqnarray}
(s\,\si'')^{2}&=&4\,s\,(\si')^{3}+
\si^{2}+(2\,\al+4\,N-2\,s)\,\si\si'\nonumber\\
&&+(\al^{2}-2\,\al\,s-4\,N\,s+s^{2})\,(\si')^{2}-4\,\si\,(\si')^{2}
\label{laguerresigmade} \end{eqnarray}
differ by a constant. (The third-order equation is equivalent to the two sides'
here
having the same derivative.) Now it is clear that if $\al$ is sufficiently
large then
$\si$ is twice continuously differntiable up to $s=0$ and $\si(0)=\si'(0)=0$.
Hence both
sides of (\ref{laguerresigmade}) vanish at $s=0$ and so the difference in
question
equals 0. Thus the identity is established for $\al$ large. But both sides of
the
identity are (for $s>0$) real-analytic for $\al>-1$ and so if they agree for
large
$\al$ they must agree for all $\al$.

Comparing (\ref{laguerresigmade}) this with (C.45) of \cite{jm} we see that
$-\si(s)$
satisfies the $\si$ version of Painlev\'{e} V with parameters
$\nu_{0}=\nu_{1}=0,\,
\nu_{2}=N$ and $\nu_{3}=N+\al$.

The boundary condition at $s=0$ for $\si$ depends, of course, on the parameter
$\la$
in (\ref{lagppdef}),
 and we write $\si(s;\la)$ instead of $\si(s)$ to display this
dependence. With the help of the Neumann series for the resolvent kernel, we
compute the
small $s$ expansion
\begin{eqnarray*}
\sigma(s;\lambda)&=& \lambda c_0 s^{\alpha+1}\left(1-{\alpha+2N\over 2+\alpha}
s + \cdots\right ) \\
&&+ \lambda^2{c_0^2\over 1+\alpha} s^{2\alpha+2}\left(1-
{(2\alpha+3)(\alpha+2N)\over (2+\alpha)^2} s + \cdots \right)\\
&&+ \lambda^3 {c_0^3\over (1+\alpha)^2}s^{3\alpha+3}\left(1+\cdots\right)\\
&& +\cdots  \end{eqnarray*}
where
\[ c_{0}={\Gamma(N+\alpha+1)\over \Gamma(N)\Gamma(\alpha+1)\Gamma(\alpha+2) }\>
.\]

 And now, as in the case of the GUE, we have a
representation
\bq P(\la_{\mbox{\footnotesize min}}>s)=\exp\left\{-\int_{0}^{s}
\fr{\si(t;1)}{t}\,dt\right\}. \label{eigenrep} \eq
There are only minor changes required in the above analysis when we take
$J=(s,\iy)$
and this leads to an analogous representation for the distribution function of
$\la_{\mbox{\footnotesize max}}$.
\subsubsection{Singular values of rectangular matrices}
If $A$ is an $N\times M$ rectangular matrix ($N\leq M$) whose entries are
independent
identically distributed complex Gaussian variables with mean 0 and variance 1
then the
$N\times N$ matrix $A\,A^{*}$ (whose eigenvalues are the squares of the
singular values
of $A$) belongs to the orthogonal polynomial ensemble associated with the
weight
function
\[ w(x)=x^{M-N}\,e^{-x/2}. \]
(See \cite{edel}, Cor. 3.1.) It follows that the distribution of the smallest
singular value of $A$ is given by the right side of (\ref{eigenrep}) with $s$
replaced by $\st{s}/2$ and, of course, $\al=M-N$. There is a similar
representation for the distribution
function of the largest singular value of $A$.

\subsection{Jacobi Kernel}
The situation here is so similar to the preceding that we shall only indicate
the main
points. For the finite $N$ Jacobi ensemble
\[ w(x)=(1-x)^{\al}(1+x)^{\beta} \]
with $\al,\,\beta>-1$ and we must take in (\ref{kernel})
\[ \vph(x)=\st{\la\, a_{N}}\> \vph_{N}(x),
\quad \psi(x)=\st{\la\, a_{N}}\> \vph_{N-1}(x)\]
where
\[ a_N={1\over N+(\alpha+\beta)/2}\>\sqrt{{N(N+\alpha)(N+\beta)(N+\alpha+\beta)
\over (2N+\alpha+\beta-1)(2N+\alpha+\beta+1)}}.\]
{}From the differentiation and recurrence
 formulas for the Jacobi polynomals we have
the differentiation formulas (\ref{diff_formulas}) with
\[m(x)=1-x^{2},\;A(x)=\al_{0}+\al_{1}\,x,\;B(x)=\be_{0},\;C(x)=\ga_{0},\]
where
\begin{eqnarray*}
\al_{0}&=& {\be^2-\al^2\over 2(2N+\al+\be)},\\
\al_{1}&=&- \left(N+\fr{\al+\be}{2}\right),\\
\be_{0}&=& \fr{2}{2N+\al+\be}\st{N(N+\al)(N+\be)(N+\al+\be)}\>
\st{\fr{2N+\al+\be+1}{2N+\al+\be-1}}\> ,\\
\ga_{0}&=&\fr{2}{2N+\al+\be}\st{N(N+\al)(N+\be)(N+\al+\be)}\>
\st{\fr{2N+\al+\be-1}{2N+\al+\be+1}}\> .
\end{eqnarray*}

Using these the interested reader could without difficulty write down the
Jacobi
analogues of the general equations (\ref{laguerre_qj})--(\ref{laguerreRjjder})
in
the Laguerre ensemble.
 We shall restrict ourselves here to the case $J=(-1,s)$, the
analogue
 of the interval $(0,s)$ in Laguerre, and find the following Jacobi analogues
of (\ref{laguerre_q0s})--(\ref{uv0s}):
\begin{eqnarray}
(1-s^{2})\,q'&=&\left(\al_{0}+\al_{1}\,s+v\right)\,q+
\left(\be_{0}+(2\al_{1}-1)\,u
\right)\,p, \label{jacq} \\
(1-s^{2})\,p'&=&\left(-\ga_{0}+(2\al_{1}+1)w\right)\,q-
\left(\al_{0}+\al_{1}\,s+
v\right)\,p, \label{jacp} \\
(1-s^{2})\,R(s)&=&2\,\left(\al_{0}+\al_{1}\,s+v\right)\,p\,q+
{\cal A} +{\cal B}, \label{jacR} \\
\left((1-s^{2})\,R(s)\right)'&=&2\,\al_{1}\,p\,q,\label{jacRder} \\
v'&=&p\,q \label{jacvder}
\end{eqnarray}
where
\[{\cal A}=\left(\be_{0}+ (2 \al_{1}-1)\, u\right)\, p^{2},\quad
{\cal B}=\left(\ga_{0}-(2 \al_{1}+1) w\right)\, q^{2}. \]
{}From (\ref{jacRder}) and (\ref{jacvder}) we deduce
\bq(1-s^{2})\,R(s)=2\,\al_{1}\,v \label{jacRv} \eq
(note that both sides here vanish when $s=-1$) and (\ref{jacq}) and
(\ref{jacp}) give
\bq (1-s^{2})\,(p\,q)'={\cal A} - {\cal B}. \label{jacpqder} \eq
Another differentiation gives
\begin{eqnarray}
(1-s^{2})\,(p\,q)''&=& -2\,\left(\al_{0}+\al_{1}\,s+v\right)\,({\cal A}+{\cal
B})
+2\, s\, ({\cal A}-{\cal B}) \nonumber \\
&&+4\,\al_{1}\,(1-s^{2})\,p^{2}\,q^{2}-\fr{4}{p\,q}\,
{\cal A}\, {\cal B}. \label{jacpqder2}
\end{eqnarray}
Now (\ref{jacRder}) and (\ref{jacRv}) show that $p\,q$ and $v$ are expressible
in terms of $R$ and its first derivative. And so, using (\ref{jacR}) and
(\ref{jacpqder}), we deduce that ${\cal A}$ and
${\cal B}$ are expressible in terms of $R$ and
its first two derivatives. Finally, using (\ref{jacpqder2}), we obtain a third
order
equation for $R$. Instead of this we write down the analogue of (\ref{lag3rd}).
If we
define now $\si(s):=(1-s^{2})\,R(s)$ then we have
\begin{eqnarray}
(1-s^{2})^{2}\,\si'''&=&(1-s^{2})^{2}\,\fr{(\si'')^{2}}{\si'}-
2\,s\,(1-s^{2})\,\si''
-2\,(1-s^{2})\,(\si')^{2} \nonumber \\
&&-2\,\left(1-\fr{2\,\al_{1}^{2}}{\si'}\right)\,\si^{2}-4\,\al_{1}\,(\al_{0}
+\al_{1}\,s)\,\si. \label{jac3rd}
\end{eqnarray}

For the boundary condition at $s=-1$ we compute the small $s+1$ expansion
to be (with an obvious notation)
\[\si(s;\la)=\la c_0(1+s)^{1+\beta}\left(1+
\fr{\beta(1-\alpha)-2N(\alpha+\beta+N)}{ 2(2+\beta)} (1+s)+\cdots\right) \]
\[+\fr{\lambda^2 c_0^2}{2 (1+\beta)} (1+s)^{2+2\beta}\left(1+
\fr{2+6\beta+3\beta^2-(3+2\beta)(2N^2+2N\alpha+2N\beta+\alpha\beta)}
{2 (2+\beta)^2}\> (1+s)+\cdots\right) \]
\[+\fr{\lambda^3 c_0^3}{ 4 (1+\beta)^2}(1+s)^{3+3\beta}\left(1+\cdots\right)
+ \cdots \]
where
\[ c_{0}= {\Gamma(N+\alpha+\beta+1)\Gamma(N+\beta+1)\over 2^\beta
\Gamma(\beta+1)
\Gamma(\beta+2)\Gamma(N)\Gamma(N+\alpha)}\,.\]
We have not been able to find a first integral for (\ref{jac3rd}), in
other words a second order equation which is analogous to
(\ref{laguerresigmade}).

\subsection{The Circular Ensemble}

\subsubsection{The partial differential equations}
If ${\cal U}(N)$ denotes the group of $N\times N$ unitary matrices, then
the finite $N$ circular ensemble
of unitary matrices (sometimes denoted CUE) is this set
${\cal U}(N)$ together with
the normalized Haar measure.
Just as for the orthogonal
polynomial hermitian matrix ensembles, the level spacing distributions are
expressed in terms of a Fredholm determinant of an integral
operator defined now on the unit circle.  All of this is
well-known and we refer the reader to either Dyson \cite{dyson62} or Mehta
\cite
{mehta_book}
for details.
\par
The integral operator for the finite $N$ CUE is
\[K\doteq \frac{\lambda}{2\pi}\, \frac{\sin \frac{N}{2}(x-y)}
{\sin \frac{1}{2}(x-y)}\,\chJ (y)
=\frac{\phi(x)\psi(y)-\psi(x)\phi(y)}{e^{ix}-e^{iy}}\,\chJ (y)\]
where
\[
\varphi(x)=\sqrt{\frac{\lambda}{2\pi}}\,e^{i\frac{N+1}{2}x},\;\;
\psi(x)=\sqrt{\frac{\lambda}{2\pi}}\,e^{-i\frac{N-1}{2}x}.\]
Thus the differentiation formulas (\ref{diff_formulas_circle}) hold with $b=i$
and
\[ m(x)=1, \ \ \ A(x)=\fr{i}{2}\, N, \ \ \ B(x)=C(x)=0 \]
and so from the considerations in Sec.~\ref{subsec:DE_circle} we deduce that
\bq R(a_{j},a_{k})=\fr{q_{j}\,p_{k}-p_{j}\,q_{k}}{e^{ia_{j}}-e^{ia_{k}}}
\qquad (j\neq k)\, , \label{Rjkcue} \eq
\begin{equation}
i\,e^{ia_{j}}\,R(a_{j},a_{j}) =i\,N\,p_{j}\,q_{j}
-\sum_{k\neq j}\,(-1)^{k}\,R(a_{j},a_{k})\,(q_{k}\,p_{j}-p_{k}\,q_{j}),
\label{Rjj_cue}\end{equation}
and
\begin{eqnarray}
{\partial q_{j}\over \partial a_{j}}&=&i\frac{N+1}{2}\,q_{j}
  -\sum_{k\neq j}\,(-1)^{k}\,R(a_{j},a_{k})\,q_{k}, \nonumber  \\
{\partial p_{j}\over \partial a_{j}}&=&-i\frac{N-1}{2}\,p_{j}
  -\sum_{k\neq j}\,(-1)^{k}\,R(a_{j},a_{k})\,p_{k}.\label{qj_pj_cue}
\end{eqnarray}

\subsubsection{Single interval case}
Now we specialize to $J=(-t,t)$ and take
$a_{1}=-t, a_{2}=t$. Write $p$, $q$ for $p_{2}$, $q_{2}$ and note that
since
$\overline{\varphi(x)}=\varphi(-x)$ and $K(x,y)$ is even we have
$p_{1}=\bar{p},\;q_{1}=\bar{q}$. Let's also write
\[R(t):=R(t,t),\;\;\tilde R(t):=R(-t,t)=R(t,-t).\]
Then (\ref{GG}) with $J=(-1,1)$ and $i=2$ gives
\begin{equation} R'\,=\,2\,\tilde R^{2}
\ \ \ \ \  \ \left(^\prime=\fr{d}{dt}\right),
  \label{gaudin2_cue}\end{equation}
(\ref{Rjj_cue}) gives
\begin{equation}
e^{it}\,R(t)=N\,p\,q-i\,\tilde R(t)\,(\bar{q}p-\bar{p}q),
\label{Rpq_cue}\end{equation}
whereas (\ref{Rjkcue}) with $j=2, k=1$ gives
\bq
q\bar{p}-p\bar{q}=2i\sin t\;\tilde R. \label{tRpq_cue}\eq
Substituting  (\ref{tRpq_cue}) into (\ref{Rpq_cue}) and using
(\ref{gaudin2_cue})
 give
\bq   e^{it}\,R(t)=N\,p\,q\,-\sin t\,R'(t).    \label{Req1_cue}\eq
Now
\[ q'=i\frac{N+1}{2}q+\tilde R\, \bar{q},\;\;p'=-i\frac{N-1}{2}p+\tilde R\,
\bar
{p}. \]
Using these and (\ref{tRpq_cue}) we get
\begin{eqnarray*}
(2i\sin t\,\tilde R)'&=&\left(i\frac{N+1}{2}q+\tilde R\,\bar{q}\right)\,
\bar{p}+q\left(i\frac{N-1}{2}\bar{p}+\tilde R
\,p\right)\\
&&-\,\left(-i\frac{N-1}{2}p+\tilde
R\,\bar{p}\right)\bar{q}-p\left(-i\frac{N+1}{
2}\bar{q}+
\tilde R\,q\right) \\
&=&iN\left(q\bar{p}-p\bar{q}\right).
\end{eqnarray*}
This and (\ref{tRpq_cue}) may be written
\begin{eqnarray*}
(\sin t\,\tilde R)'&=&N\,\Re\,(q\,\bar{p}),\\
\sin t\,\tilde R&=&\Im\,(q\,\bar{p}),
\end{eqnarray*}
and (\ref{Req1_cue}) may be written
\[ e^{it}\,R(t)\,+\,\sin t\,R'(t)\,=\,N\,p\,q.\]
Thus (taking the square of the absolute value of both sides)
\[ \left(\cos t\,R(t)+\sin t\,R'(t)\right)^{2}\,+\,\sin^{2}\!t\,R(t)^{2}
=(\sin t\,\tilde R)'\,^{2}\,+\,N^{2}\,\sin^{2}\!t\,\tilde R^{2}.\]
Using \[\tilde R^{2}=\frac{1}{2}R',\ \ \tilde R'\tilde R=\frac{1}{4}R'', \ \
\tilde R'^{\,2}=\frac{1}{8}
\frac{R''^{\,2}}{R'},\]
which follow from (\ref{gaudin2_cue}), we get the second order equation
\begin{eqnarray}
&& R(t)^{2}+2\,\sin t\,\cos t\,R(t)\,R'(t)+\sin^{2}\!t\,R'(t)^{2}\nonumber \\
&=& \frac{1}{2}\left(\frac{1}{4}\sin^{2}\!t\frac{R''(t)^{2}}{R'(t)}
+\sin t\,\cos t\,R''(t)+(\cos^{2}\!t+N^{2}\sin^{2}\!t)\,R'(t)\right).
\label{Rde_cue}\end{eqnarray}
\par
{}From the Neumann expansion of the resolvent kernel
$R(t;\la)=R(t)$  we
 obtain the expansion
\bq
R(t,\la)=
\rho_0 + 2 \rho_0^2 t + 4 \rho_0^3 t^2 +\fr{2}{9} (1-N^2+36\rho_0^2)\rho_0^2
t^3
+ \cdots \label{R_smallt}\eq
where
\[ \rho_0 = \fr{\la N}{2\pi}. \]
If we denote by $E_{N}(0;s)$ the probability that an interval (of the unit
circle)
of length $s$ contains no eigenvalues (modifying here the notation of
(\ref{detform})),
then
\[ E_{N}(0;s) = \exp\left\{ -2 \int_0^t R(x,1)\, dx \right\}
\ \ \ (s=2t). \]
Using the expansion
 (\ref{R_smallt})
with $\la=1$ (and additional terms computed from
the differential equation (\ref{Rde_cue})) we find that
\begin{eqnarray*}
 E_{N}(0;\fr{2\pi}{N} s)&=& 1 - s +\fr{\pi^2}{36}\left(1-\fr{1}{N^2}\right) s^4
-\fr{\pi^4}{675}\left(1-\fr{5}{2N^2}+\fr{3}{2N^4}\right) s^6 \\
&& \ \
+\fr{\pi^6}{17640}\left(1-\fr{14}{3N^2}+\fr{7}{N^4}-\fr{10}{3N^6}\right)s
^8 + \cdots
\ \ \ (s\ra 0)\end{eqnarray*}
where we have replaced $s$ by $\fr{2\pi}{N} s$ so that  the $N\ra \iy$
limit  is clear. This converges uniformly for all $N$ and bounded $s$.
Observe that the corrections to the limiting coefficients
are ${\rm O}(1/N^2)$ as $N\ra \iy$.

%%%%%%%%%%%%%%%%%%%%%%%%%%%%%%%%%%%%%%%%%%%%%%%%%%%%%%%%%%%%%%%%%%%%%%%%%%%%
\section{Generalizations of Hermite, Laguerre and Jacobi}
\label{sec:generalN}
In this final section we shall show that there are differentiation
formulas of the form (\ref{diff_formulas}) for very general orthogonal
polynomial ensembles, and that if the weight function is the standard
Hermite, Laguerre, or Jacobi weight function multiplied by the exponential
of an arbitrary polynomial then the coefficients $m(x)$, $A(x)$,
$B(x)$ and $C(x)$ in (\ref{diff_formulas}) are themselves polynomials.
Some, but not all, of our derivation can be found in the
orthogonal polynomial literature \cite{bauldry,bonan_clark} but our
presentation
will be self-contained.
\par
Throughout, we shall write our weight function as
\[ w(x)=e^{-V(x)}. \] As stated in the Introduction,
the polynomials orthonormal  with respect to $w(x)$ are denoted
$p_k(x)$ ($k=0,1,\cdots$), and we set $\varphi_k(x)=p_k(x) w(x)^{1/2}$
so that $\{\varphi_k\}$ is orthonormal with respect to Lebesgue
measure.  The underlying domain ${\cal D}$ of all these functions
we take to be a finite or infinite interval.  We are interested
in differentiation formulas (\ref{diff_formulas}) when, up to constant factors,
\bq \varphi(x)=\varphi_N(x),\ \ \ \psi(x)=\varphi_{N-1}(x).
\label{phi_psi}
\eq
\par
It is well-known that if $k_N$ denotes the highest coefficient
in $p_N(x)$, and if
\[ a_N=\fr{k_{N-1}}{k_N}\> , \]
then there is a recursion formula
\bq
x p_N(x) = a_{N+1} p_{N+1}(x) + b_N p_N(x) + a_N p_{N-1}(x)
\label{OP_recursion}\eq
as well as the Christoffel-Darboux formula
\bq
\sum_{k=0}^{N-1} p_k(x) p_k(y) = a_N \,  { p_N(x) p_{N-1}(y) - p_{N-1}(x)
p_N(y)
\over x-y }\; . \label{christoffel} \eq
(See Chap.~10 in \cite{erdelyi}.)
\par
We shall always assume that our weight function satisfies
\[ x^k w(x) \ \ \mbox{ is bounded for each}\ k=0,1,\cdots \]
and that $V(x)$ is continuously differentiable in the interior of ${\cal D}$.
And we assume that
\bq \lim_{x\ra \pl{\cal D}} w(x) = 0 \label{edge_condition}\eq
although this will be relaxed later.  We define \cite{bauldry,bonan_clark}
\begin{eqnarray*}
 U(x,y)& =& \fr{V'(x)-V'(y)}{x-y}\; , \\
A_N(x)&= &a_N \int_{{\cal D}} \varphi_N(y) \varphi_{N-1}(y) U(x,y)\, dy, \ \ \
B_N(x)= a_N \int_{{\cal D}} \varphi_N(y)^2 U(x,y)\, dy\, .
\end{eqnarray*}
\medskip\bigbreak\bigbreak \par
\noindent {\bf Proposition}: {\it If (\ref{edge_condition}) holds and $\varphi$
and
$\psi$ are given by (\ref{phi_psi}) then we have (\ref{diff_formulas})
with}
\begin{eqnarray*}
m(x)&=&1,\ \ \ \ \ \ \    A(x)= - A_N(x) - \fr{1}{2} V'(x), \\
B(x)&=& B_N(x),\ \ \   C(x)= \fr{a_N}{a_{N-1}} B_{N-1}(x).
\end{eqnarray*}
\smallskip
\noindent {\it Proof}. If
\[ \tilde K(x,y):= \sum_{k=0}^{N-1} p_k(x) p_k(y) \]
then for any polynomial $\pi(x)$ of degree at most $N-1$
\[ \pi(x) = \int \tilde K(x,y) \pi(y) w(y)\, dy. \]
Apply this to $\pi(x)=p_N'(x)$ and integrate by parts, using
 (\ref{edge_condition}) to eliminate any boundary terms.  We obtain
\[ p_N'(x)=-\int \fr{\pl \tilde K(x,y)}{\pl y} p_N(y) w(y)\, dy
+ \int \tilde K(x,y) V'(y) p_N(x) w(y)\, dy .\]
\par
Now both $\pl\tilde K/\pl y$ and $\tilde K$, as polynomials in $y$,
are both orthogonal (with respect to $w$) to $p_N(y)$ since they
have degree at  most $N-1$.  It follows that the first integral above
vanishes and that we can write the resulting identity as
\begin{eqnarray*}
p_N'(x)&=&\int \tilde K(x,y)\left( V'(y)-V'(x)\right) p_N(y) w(y)\, dy \\
&=& - a_N \int \left[ p_N(x) p_{N-1}(y)-p_{N-1}(x) p_N(y)\right]
p_N(y) U(x,y) w(y)\, dy
\end{eqnarray*}
by (\ref{christoffel}) and the definition of $U(x,y)$.  We have
shown
\bq p_N'(x) = - A_N(x) p_N(x) + B_N(x) p_{N-1}(x).
\label{p_N'}\eq
\par
It follows from this, of course, that
\[ p_{N-1}'(x) = - A_{N-1}(x) p_{N-1}(x) + B_{N-1}(x) p_{N-2}(x). \]
We use the recursion formula (\ref{OP_recursion}) (which holds also
if each $p_N$ is replaced by $\varphi_N$) and find that this is equal
to
 \begin{eqnarray*}
&& - \int \varphi_{N-1}(y)\left[ y \varphi_{N-1}(y) - a_n \varphi_N(y)
- b_{N-1} \varphi_{N-1}(y)\right] U(x,y)\, dy \; p_{N-1}(x) \\
&& + \int \varphi_{N-1}^2(y) U(x,y)\, dy \; \left[ x p_{N-1}(x) - a_N p_N(x)
- b_{N-1} p_{N-1}(x)\right] \\
&=&\int \varphi_{N-1}^2(y) (x-y) U(x,y)\, dy\; p_{N-1}(x)
+ A_N(x) p_{N-1}(x) - \fr{a_N}{a_{N-1}} B_{N-1}(x) p_N(x).
 \end{eqnarray*}
The last integral on the right side equals
\[ \int p_{N-1}^2(y) \left(V'(x)-V'(y)\right)w(y)\, dy =
V'(x) + \int p_{N-1}^2(y) w'(y)\, dy .\]
The last integral vanishes, as we see by integrating by parts and
noting that $p_{N-1}'$ is orthogonal to $p_{N-1}$.  Thus we
have shown
\bq
p_{N-1}'(x) =\left[ A_N(x) + V'(x)\right] p_{N-1}(x) -
\fr{a_N}{a_{N-1}} B_{N-1}(x) p_N(x). \label{p_N-1'}\eq
\par
The statement of the proposition now follows from (\ref{p_N'}) and
(\ref{p_N-1'}) if we use the fact that
\[ \varphi_N'=p_N' w^{1/2} + \fr{1}{2} \fr{w'}{w} p_N w^{1/2}, \]
and similarly for $\varphi_{N-1}'$. \qed\par
\medskip
\noindent {\it Remark\/}.  The assumption (\ref{edge_condition}) is not just a
technical
requirement.  The conclusion of the proposition is false without it.
(Consider, for example, the Legendre polynomials on $(-1,1)$, where
$V(x)=0$ and the conclusion of the proposition reads $\varphi_n'(x)=0$.)\ \
Nevertheless, we shall be able to handle some cases where
 (\ref{edge_condition}) fails.
\par \medskip
\noindent {\it Example 1 (generalized Hermite)\/}.
Here $V(x)$ is a polynomial of even degree (at least 2) with positive
leading coefficient and ${\cal D}=(-\iy,\iy)$.  The conclusion of the
proposition holds and so we have (\ref{diff_formulas}) with
$m(x)=1$, with $A(x)$ a polynomial of degree at most $\deg V-1$,
and with $B(x)$ and $C(x)$ polynomials of degree at most $\deg V-2$.
\par
\medskip\bigbreak\bigbreak
\noindent {\it Example 2 (generalized Laguerre)\/}.  Here
\[ w(x) = x^\al e^{-W(x)} \]
where $\al>-1$ and $W$ is a polynomial of degree at least $1$ with positive
leading coefficient, and ${\cal D}=(0,\iy)$.  In this case
\bq
U(x,y)= \fr{\al}{xy} + \fr{W'(x)-W'(y)}{x-y}\> . \label{laguerre_U} \eq
Now (\ref{edge_condition}) is satisfied if $\al>0$ and the proposition
tells us that in this case we have (\ref{diff_formulas}) with
\begin{eqnarray}
m(x)&=&x, \ \ \ \ \ A(x)= -x A_N(x) + \fr{\al}{2} - \fr{x}{2} W'(x),
\nonumber\\
B(x)&=& x B_N(x), \ \ \ C(x)= \fr{a_N}{a_{N-1}}
x B_{N-1}(x)\, ;\label{laguerre_diff}
\end{eqnarray}
now $A(x)$ is a polynomial of degree at most $\deg W$ while $B(x)$
and $C(x)$ are polynomials of degree at most $\deg W -1 $.
\par
To extend this to all $\al>-1$ we see that there are problems
in the integrals defining $A_N(x)$ and $B_N(x)$ arising from the
term $\al/xy$ in (\ref{laguerre_U}).  The contribution of this term
to the integral defining $A_N(x)$, say, equals (we assume now
$\al>0$)
\[ a_N\fr{\al}{x}\int_0^\iy p_N(y) p_{N-1}(y) e^{-W(y)} y^{\al-1}\, dy\, . \]
Integration by parts shows that this equals
\[ -\fr{a_N}{x} \int_0^\iy \left( p_N(y) p_{N-1}(y) e^{-W(y)}\right)^\prime
y^\al \, dy\, . \]
This expression is well-defined for all $\al>-1$ and in fact represents
a function of $\al$ which is real-analytic there.  (The coefficients
of the $p_N$ are clearly real-analytic functions of $\al$.)\ \
\par
This argument shows that both sides of (\ref{diff_formulas}), with
the coefficient polynomials given by (\ref{laguerre_diff}),
are (or extend to be) real-analytic for $\al>-1$.  Since they agree
for $\al>0$ they must also agree for $\al>-1$.
\par
\medskip\bigbreak\bigbreak
\noindent {\it Example 3 (generalized Jacobi)\/}. Here
\[ w(x) = (1-x)^\al (1+x)^\be e^{-W(x)} \]
where $\al,\be>-1$ and $W$ is a polynomial, ${\cal D}=(-1,1)$.
In this case
\[ U(x,y)={\al(1+x)(1+y)+\be (1-x)(1-y)\over (1-x^2)(1-y^2)}+
{W'(x)-W'(y)\over x-y} \]
and the proposition tells us that for $\al, \be>0$ we have
(\ref{diff_formulas}) with
\begin{eqnarray*}
m(x)&=& 1-x^2, \ \ \ A(x)=-(1-x^2) A_N(x) - \fr{\al}{2}(1+x) +
\fr{\be}{2} (1-x) - \fr{1-x^2}{2} W'(x), \\
B(x)&=& (1-x^2) B_N(x), \ \ \ C(x)=\fr{a_N}{a_{N-1}}(1-x^2) B_{N-1}(x);
\end{eqnarray*}
now $A(x)$ is a polynomial of degree at most $\deg W +1$ while
$B(x)$ and $C(x)$ are polynomials of degree at most $\deg W$.  The identity
can be extended to $\al,\be>-1$ as in Example 2.  (It is convenient
to express the integrals defining $A_N(x)$ and $B_N(x)$ as sums of
integrals by using a representation
\[ 1 = u(x) + \left(1-u(x)\right) \]
where $u=1$ in a neighborhood of $x=-1$ and $u=0$ in a neighborhood
of $x=1$; this separates the difficulties at the two end-points.  The details
are left to the reader.)\ \
\medskip
\par \noindent
{\it Remark\/}. It is clear that the last examples  can be generalized to any
weight
function of the form
\[ \prod |x-a_{i}|^{\al_{i}}e^{-W(x)} \]
where $W(x)$ is a polynomial and for each $a_{i}$
which is in the closure of ${\cal D}$
we have $\al_i>-1$.

\acknowledgments
Over the past eighteen months the authors have benefited from
an ongoing  correspondence with Professor Freeman J. Dyson regarding
random matrices.  It is a pleasure to acknowledge this and,  as well,
the support provided by the National Science Foundation through
grants DMS--9001794, DMS--9303413, and   DMS--9216203.

\end{document}